\documentclass[12pt,a4paper]{article}
\pdfoutput=1
\usepackage{jheppub}
\usepackage{empheq}
\usepackage{epstopdf}
\usepackage{graphicx}
\usepackage{epsfig}
\usepackage{dcolumn}  
\usepackage{bm}    
\usepackage{amssymb} 

\usepackage{amsmath,bm}
\usepackage{amsfonts}    
\usepackage{slashed}  
\usepackage[mathscr]{euscript}
\usepackage{epsfig}
\usepackage[position=t,singlelinecheck=off]{subfig}
\usepackage{caption}
\allowdisplaybreaks[1]

\title{Partition function of ${\cal N}=2^*$ SYM on a large four-sphere} 

\author{Timothy J. Hollowood and S. Prem Kumar} 
\affiliation{Department of Physics, Swansea University,\\Singleton Park, Swansea SA2 8PP, UK.}
\emailAdd{t.hollowood, s.p.kumar@swansea.ac.uk}
\abstract{We examine the partition function of ${\cal N}=2^*$ supersymmetric $SU(N)$ Yang-Mills theory on the four-sphere in the large radius limit. We point out that the large radius partition function, at fixed $N$, is computed  by saddle-points lying on walls of marginal stability on the Coulomb branch of the theory on ${\mathbb R}^4$. For $N$ an even (odd) integer and $\theta_{\rm YM}=0\, (\pi)$, these include a point of maximal degeneration of the Donagi-Witten curve to a torus where BPS dyons with electric charge $\left[\frac{N}{2}\right]$ become massless.  We argue
that the dyon singularity is the lone saddle-point in the $SU(2)$ theory, while for $SU(N)$ with $N>2$, we characterize potentially competing saddle-points by obtaining the  relations between the Seiberg-Witten periods at such points. Using 
Nekrasov's instanton partition function, we solve for the maximally degenerate saddle-point and obtain its free energy as a function of $g_{\rm YM}$ and $N$, and show that the results are ``large-$N$ exact''.  In the large-$N$ theory our results provide analytical expressions for the periods/eigenvalues at the maximally degenerate saddle-point,  precisely matching previously known formulae following from the correspondence between ${\cal N}=2^*$ theory and the elliptic Calogero-Moser integrable model. The maximally singular point ceases to be a saddle-point of the partition function above a critical value of the coupling, in agreement with the recent findings of Russo and Zarembo.
}

\begin{document}
 \maketitle 

\def\Xint#1{\mathchoice
   {\XXint\displaystyle\textstyle{#1}}%
   {\XXint\textstyle\scriptstyle{#1}}%
   {\XXint\scriptstyle\scriptscriptstyle{#1}}%
   {\XXint\scriptscriptstyle\scriptscriptstyle{#1}}%
   \!\int}
\def\XXint#1#2#3{{\setbox0=\hbox{$#1{#2#3}{\int}$}
     \vcenter{\hbox{$#2#3$}}\kern-.5\wd0}}
\def\ddashint{\Xint=}
\def\dashint{\Xint-}
\def\tq{\tilde q}
\def\be{\begin{equation}}
\def\ee{\end{equation}}
\def\bea{\begin{eqnarray}}
\def\eea{\end{eqnarray}}
\def\nn{\nonumber}
\def\pd{\partial}
\def\ttau{\tilde\tau}
\def\tG{\tilde G}
\def\Re{R\'{e}nyi }
\def\l1{{\text{1-loop}}}
\def\uy{u_y}
\def\ur{u_R}
\def\o{\mathcal{O}}
\def\Cl{{{cl}}}
\def\bz{{\bar{z}}}
\def\by{{\bar{y}}}
\def\bX{\bar{X}}
\def\im{{\text{Im}}}
\def\re{{\text{Re}}}
\def\cn{{\text{cn}}}
\def\sn{{\text{sn}}}
\def\dn{{\text{dn}}}
\def\K{\mathbf{K}}
\def\n1{\Bigg|_{n=1}}
\def\fin{{\text{finite}}}
\def\R{{\mathscr{R}}}
\def\one{{(1)}}
\def\zero{{(0)}}
\def\n{{(n)}}
\def\tr{\text{Tr}}
\def\T{\mathcal{T}}
\def\TT{\tilde{\mathcal{T}}}
\def\O{\mathcal{O}}
\def\cN{\mathcal{N}}
\def\P{\Phi}
\def\csch{{\text{cosech}}}
\def\W{{\tilde{W}}}
\def\T{{\tilde{T}}}
\def\by{\bar{y}}
\newcommand*\xbar[1]{%
  \hbox{%
    \vbox{%
      \hrule height 0.5pt 
      \kern0.5ex
      \hbox{%
        \kern-0.1em
        \ensuremath{#1}%
        \kern-0.1em
      }%
    }%
  }%
} 
\section{Introduction and summary}
Localization techniques have emerged as a powerful and elegant tool for extracting nonperturbative information on quantum field theories in various dimensions. In particular, Pestun's work \cite{pestun} provides a remarkable and concrete  formulation of  the partition function of supersymmetric (SUSY) gauge theories on spheres, in terms of ordinary (matrix) integrals. This formulation allows the exact computation of field theoretic observables such as supersymmetric Wilson loops \cite{pestun} which could then be compared and matched with corresponding results for large-$N$ gauge theories with holographic supergravity duals \cite{drukkergross,esz} e.g. the ${\cal N}=4$ SUSY Yang-Mills (SYM) theory in four dimensions. The matrix models for ${\cal N}=2$ theories with flavours following from Pestun's work,  were further explored in the large-$N$ limit at strong coupling \cite{Passerini:2011fe, Fraser:2011qa} to deduce aspects of putative string duals of such theories.

In this paper, motivated by the works of Russo and Zarembo  \cite{zaremborusso1, zaremborusso2, zaremborusso3}, we investigate certain aspects of the partition function of $SU(N)$, ${\cal N}=2$ SYM with one massive adjoint hypermultiplet, on the four-sphere. This theory, also known as ${\cal N}=2^*$ SYM, is the ${\cal N}=2$ supersymmetric mass deformation of ${\cal N}=4$ SYM. In references \cite{zaremborusso1, zaremborusso2, zaremborusso3} it was found that the large-$N$ partition function of ${\cal N}=2^*$ theory on $S^4$, in the large radius limit,  undergoes an infinite sequence of quantum phase transitions with increasing 't Hooft coupling $\lambda$. 

One of several intriguing aspects of this picture is that the low-$\lambda$ phase of \cite{zaremborusso1, zaremborusso2}, (for $0<\lambda\leq \lambda_c\approx 35.45$) has exactly calculable condensates which coincide precisely with the exact results (obtained sometime ago in \cite{ Dorey:1999sj, Dorey:2002ad, mm}) for a specific maximally degenerate point on the Coulomb branch of ${\cal N}=2^*$ theory on ${\mathbb R}^4$.  At such a  point the Seiberg-Witten curve for the theory \cite{sw, sw2,dw} undergoes maximal degeneration due to the appearance of $N-1$ massless, mutually local BPS states. 
The total number of maximally degenerate vacua of ${\cal N}=2^*$ theory is given by a sum over all the divisors of $N$ (for the $SU(N)$ theory). We are immediately presented with a potential puzzle: which one of these special points is picked out as a saddle-point of the partition function and why? This question was the original motivation for our work. 

We answer the question by first noting that in the limit of large radius, regardless of $N$, Pestun's partition sum is determined by the critical points of the real part of the ${\cal N}=2$ prepotential evaluated on configurations with purely imaginary Seiberg-Witten periods \cite{sw}. Localisation of the partition function onto constant configurations yields an ordinary multi-dimensional integral over the imaginary slice of the space of $(N-1)$ independent periods $\{a_j\}$ (${j=1\ldots N}$). We point out that saddle-points lying on this integration contour must also have purely imaginary dual periods $\{a_{D\,j}\}$. When the phases of the  Seiberg-Witten periods and dual periods are aligned we encounter a wall of marginal stability \cite{sw2}. Therefore, saddle-points contributing to the large volume partition function may be viewed as the points of intersection of the marginal stability wall with the imaginary slice/contour selected by Pestun's formulation.

Working at fixed $N$, for generic values of the microscopic (UV) coupling constant $g_{\rm YM}$ and vacuum angle $\theta_{\rm YM}$, we find that the critical points on the contour described above are {\em not}  related to singular points on the Coulomb branch of the theory on ${\mathbb R}^4$. However, when $\theta_{\rm YM}=0$ with $N$ an even integer and $\theta_{\rm YM}=\pi$ for odd $N$, one of the maximally singular points lands on this contour and is also a saddle-point. In particular, at this point the massless BPS dyons each carry an electric charge $\left[\frac{N}{2}\right]$ under one distinct abelian factor on the Coulomb branch. In the large-$N$ limit this statement applies for any $\theta_{\rm YM}$ since the effect of the vacuum angle effectively scales to zero in the strict large-$N$ limit. Put slightly differently, it is well understood \cite{Dorey:1999sj, Dorey:2000fc, Aharony:2000nt} that at the maximally singular points without massless electric hypermultiplet states i.e. those that are relevant for this paper, low energy observables in ${\cal N}=2^*$ SYM depend only on the combination $\tilde \tau\equiv (\tau +k)/N$ where $\tau \equiv 4\pi i/g^2_{\rm YM}\,+\,\theta_{\rm YM}/2\pi$ and $k\,=\,0,1,2 \ldots N-1$. For such points, the dependence on $\theta_{\rm YM}$ vanishes in the limit $N\to \infty$ and the vacuum with $\frac{k}{N} \to \frac{1}{2}$ is picked out at large-$N$ as the saddle-point.

We establish the picture above by direct examination of the ${\cal N}=2^*$ prepotential which also shows that for the special situations with $\theta_{\rm YM}=0$ and $\pi$, the partition sum can have additional saddle-points which are not points of maximal degeneration. Instead, at these additional points, while a subset of the cycles are degenerate, the remaining satisfy saddle-point conditions involving linear combinations of periods with non-zero intersection numbers. This suggests a relation to Argyres-Douglas type singularities \cite{Argyres:1995jj} as  has been found recently in theories with flavours \cite{russo, Russo:2015vva}. For the $SU(2)$  ${\cal N}=2^*$ theory we provide strong evidence that the dyon singularity (which is trivially a maximal degeneration point) is the only saddle-point of the partition function on $S^4$ (when $\theta_{\rm YM}=0$). In a certain sense which we make precise, instanton contributions preclude the possibility of an additional saddle-point, confirming the expectations of \cite{russo}.

A novel aspect of our work is that for any fixed $N$  (and large $S^4$ radius) we are able to solve exactly for the maximally degenerate saddle-point  utilising the direct relationship between Pestun's partition function and  Nekrasov's instanton partition function for the ${\cal N}=2^*$ theory on the so-called $\Omega$-background 
\cite{pestun, Nekrasov:2002qd, Nekrasov:2003rj}.
The $\Omega$-deformation parameters are set by the inverse radius and in the limit of large radius, Nekrasov's partition function is dominated by a saddle-point. The saddle-point conditions in this language, as expected, pick out points on the marginal stability wall with purely imaginary periods. The point of maximal degeneration can be characterised in terms of a complex analytic function with two branch cuts  that are glued together in a certain way. Such saddle-point equations have previously appeared in  a closely related physical context, namely, in the description of the holomorphic sector of vacua of ${\cal N}=1^*$ theory using Dijkgraaf-Vafa matrix models \cite{mm,DV, kkn}. Recognizing the connection between the degenerate Donagi-Witten curve (a torus with complex structure parameter $\ttau$) and the Riemann surface picked out by the saddle-point equations we employ a unformization map to solve for the saddle-point and obtain the exact values of the condensates. These match previously known formulae obtained by other methods \cite{Dorey:2002ad} involving the correspondence between ${\cal N}=2$ gauge theories and integrable systems. The saddle-point equation for the Nekrasov partition function at the maximally singular point also makes it manifestly clear that all nontrivial dependence on  $N$ enters through the combination $\lambda=g^2_{\rm YM}N$ even when $N$ is fixed. This property of ``large-$N$ exactness'' of physical observables at maximally singular points has also been understood in the context of ${\cal N}=1^*$ vacua wherein planar graphs of the Dijkraaf-Vafa matrix model completely characterise such points \cite{mm}.

In the large-$N$ limit, we reproduce the results of \cite{zaremborusso1} and in particular, we observe that beyond  a critical value $(\lambda_c\approx 35.45)$ of the 't Hooft coupling, the Seiberg-Witten periods at the point of maximal degeneration move off the imaginary slice so that this is no longer a saddle-point. Beyond this value of the 't Hooft coupling, the partition function is computed by a different critical point as argued in \cite{zaremborusso1, zaremborusso2}. Our analysis indicates that with the exception of the $SU(2)$ theory such a phenomenon should also occur for theories at fixed $N$: beyond a certain critical value of the gauge coupling, $\lambda_c(N) > \lambda_c(N\to \infty)\simeq 35.45$, the point of maximal degeneration should cease to be a saddle-point. From the viewpoint of Seiberg-Witten theory, this  occurs when the maximally degenerate saddle point approaches another singular (non-maximal) point where one or more  massless electric hypermultiplets appear. This cannot happen for the $SU(2)$ theory since the singular points are trivially maximal and points of maximal degeneration in ${\cal N}=2^*$ theory cannot collide.
Formally we may say that for the $SU(2)$ case, $\lambda_c(2)\to \infty$. 
 
Finally, one of the most intriguing aspects of the large-$N$  partition function is that at strong coupling it appears to be computed by a particularly simple configuration characterised by the Wigner semicircle distribution of eigenvalues/periods \cite{zaremborusso1, zaremborusso2}. We point out that maximally degenerate vacua of ${\cal N}=2^*$ SYM at large-$N$ do not have the correct strong coupling behaviour to reproduce the scaling of condensates with $\lambda$ required by the Wigner distribution. 

For the sake of clarity we list the central ideas and outcomes of the analysis presented in this paper:
\begin{itemize}
\item{Making use of the large radius limit (as opposed to the large-$N$ limit) to localise the partition function on to saddle points. This has also been pointed out in other related works, notably \cite{russo}.}
\item{Employing Nekrasov's instanton ``matrix model'' functional to understand the relevant saddle points and calculate the free energies at fixed $N$.}
\item{The special role played by one of the large number of maximally singular points on the Coulomb branch of ${\cal N}=2^*$ theory.}
\item{Calculation of observables in the low-$\lambda$ saddle point for fixed $N$, as exact functions of the gauge coupling using the Nekrasov functional. }
\item{Clarification of certain aspects of the quantum phase transitions studied in earlier works \cite{zaremborusso1, zaremborusso2}, and  their manifestation in the theories at finite $N$, at large radius.}
\end{itemize}

The organisation of the paper is as follows: Section 2 commences with some basic background on ${\cal N}=2^*$ theory, the general features of the large volume limit of the partition function on $S^4$ and the connection to points of marginal stability. We then study the criteria satisfied by the Seiberg-Witten periods at the saddle-points and their connection to singular points on the Coulomb branch. The saddle point(s) of the $SU(2)$ theory are investigated in detail and the general criteria laid out for $SU(N)$. In Section 3 we review the essential aspects of Nekrasov's instanton partition function in the large volume limit and extract the saddle-point conditions relevant for the Pestun partition sum on $S^4$. We then present the detailed solution for the maximally degenerate saddle point for any $N$ and examine its features as a function the gauge coupling. Section 4 makes contact with the large-$N$ investigations of Russo and Zarembo. We conclude with a discussion of open questions and future directions. A synopsis of essential properties of elliptic functions and modular forms is presented in an appendix.

\section{${\cal N}=2^*$ theory on $S^4$}

${\cal N}=2^*$ supersymmetric (SUSY) gauge theory is the ${\cal N}=2$ SUSY preserving mass deformation of ${\cal N}=4$ SYM. It can be viewed as an 
${\cal N}=2$ vector multiplet coupled to a massive adjoint hypermultiplet. The lowest component of the ${\cal N}=2$ vector multiplet is an adjoint scalar field $\Phi$. For the theory with $SU(N)$ gauge group on ${\mathbb R}^4$ and at weak coupling, the VEVs of the eigenvalues of $\Phi$ parametrize the Coulomb branch moduli space,
\be
\Phi\,=\,{\rm diag}\left(\hat a_1,\,\hat a_2\,,\ldots \hat a_N\right)\,,\qquad \sum_{i=1}^N \hat a_i\,=\,0\,.
\ee
The effective theory on the Coulomb branch \cite{sw} is determined by the Donagi-Witten curve \cite{dw}. At a generic point on the Coulomb branch moduli space on ${\mathbb R}^4$, the Donagi-Witten curve corresponds to a Riemann surface of genus $N$ which is a branched $N$-fold cover of the  torus with complex structure parameter given by the coupling constant of the parent ${\cal N}=4$ theory
\be
\tau\,=\, \frac{4\pi i}{g^2_{\rm YM}} \,+\,\frac{\theta_{\rm YM}}{2\pi}\,.
\ee 
The Coulomb branch moduli space has special points where the Donagi-Witten curve undergoes maximal degeneration to a {\em genus one} Riemann surface\footnote{In pure ${\cal N}=2$ SYM, the Seiberg-Witten curve has genus $N-1$ and can maximally degenerate to genus zero.}. The  points of maximal degeneration on the Coulomb branch moduli space are special, in that they are in one-to-one correspondence with {\em massive vacua} of ${\cal N}=1^*$ SYM theory obtained by the ${\cal N}=1$ SUSY mass deformation of the ${\cal N}=2^*$ theory. These points which we sometimes refer to as  ``${\cal N}=1^*$ points'' will play an important role in our work below.

When the theory is formulated on $S^4$, the Coulomb branch moduli space is lifted due to the conformal coupling of the adjoint scalar fields to the curvature of the $S^4$, and the zero modes of the adjoint scalar must be integrated over as a consequence of the finite volume. Furthermore, the realisation of ${\cal N}=2$ supersymmetry on $S^4$ requires additional terms in the microscopic Lagrangian.
The supersymmetric partition function for the ${\cal N}=2^*$ theory on the four-sphere of radius $R$  is known to localize onto constant configurations and the corresponding matrix integral was deduced by Pestun \cite{pestun}.  

\subsection{Relation to Nekrasov's partition function}

Pestun's formulation of the partition function for ${\cal N}=2$ theories on $S^4$ is intimately related to Nekrasov's ${\cal N}=2$ instanton partition function on  the so-called $\Omega$-deformation of ${\mathbb R}^4$ \cite{pestun, Nekrasov:2002qd, Nekrasov:2003rj} . The connection between the instanton partition function on the $\Omega$-background
 and Pestun's partition function on $S^4$ requires the identification of the $\Omega$-deformation parameters $\epsilon_1,\epsilon_2$ with the inverse radius of $S^4$:
 \be
 \epsilon_1\,=\,\epsilon_2\,=\,R^{-1}\,,
 \ee
 so that
 \be
 {\cal Z}_{S^4}\,=\,\int d^{N-1}a\prod_{i< j}(a_i-a_j)^2 \,\left|
 {\cal Z}_{\rm Nekrasov}(ia,\,R^{-1},\,R^{-1}, iM)\right|^2\,.\label{zs4}
 \ee
$M$ is the mass of the adjoint hypermultiplet and the $\{a_i\}$ are  $N-1$ independent, real variables, related to eigenvalues of the zero mode of the adjoint scalar in the ${\cal N}=2$ vector multiplet:
\be
\hat a_j\,=\,i a_j\,,\qquad \sum_{j=1}^N a_j=0\,.
\ee
An important aspect of Nekrasov's instanton partition function is that it includes classical, one-loop and so-called instanton pieces, all at once:
\be
{\cal Z}_{\rm Nekrasov}\,=\,{\cal Z}_{\rm cl}\,{\cal Z}_{\rm 1-loop}\,
{\cal Z}_{\rm inst}\,.
\ee
 In this sense it is somewhat artificial to split the partition function on $S^4$ into perturbative and non-perturbative contributions. Such a split really depends on the appropriate duality frame in the low energy effective theory on the Coulomb branch of ${\cal N}=2$ gauge theory. We will be interested in the limit of large $S^4$ radius, or equivalently, large hypermultiplet mass which has received attention in the recent works \cite{zaremborusso1} and \cite{zaremborusso2}. From the viewpoint of Nekrasov's partition function, the large radius limit is particularly interesting since the instanton partition function is then directly given by the Seiberg-Witten prepotential for the low-energy effective theory on the Coulomb branch on ${\mathbb R}^4$:
 \be
 {\cal Z}_{\rm Nekrasov}\left(ia,R^{-1},R^{-1}, iM \right)\big|_{R^{-1}\to 0}\,\to\, \exp\left(-R^2\,{\cal F}(ia,\,iM, \,i\tau)\right)\,.
 \ee
Here ${\cal F}$ denotes the Seiberg-Witten prepotential, encapsulating classical, one-loop and all instanton corrections at the point on the Coulomb branch labelled by the coordinates $\{i a_j\}$. For the purpose of this paper $\cal F$ can be identified with the leading contribution at large $R$. Subleading terms in the large $R$ expansion correspond to a series of gravitational couplings, which will not be relevant for our discussion.Since the exponent of the instanton partition function  scales as $R^2$, the measure factor in eq.\eqref{zs4} is subleading  for large $R$, and the partition function can be evaluated on the saddle-point(s) of the integrand of
\be
{\cal Z}_{S^4} \sim \int d^{N-1}a  \,\exp\left[-R^2\left\{{\cal F}\left(ia,\,iM,\,i\tau \right)\,+\,\overline{{\cal F}\left( ia,\, iM,\,i\tau \right)}\right\}\right]\,.\label{zlargevol}
\ee
The saddle-point conditions are non-trivial,
\be
\frac{\partial {\cal F}}{\partial a_j}\,+\,
\frac{\partial \overline{{\cal F}}}{\partial a_j}\,=\,0\,,
\qquad\qquad {j=1,2,\ldots N}\,,\label{saddleF}
\ee
and must be interpreted with care, since the prepotential ${\cal F}$ is a multivalued function with branch cuts. Recalling the definition of the dual periods in Seiberg-Witten theory, following the conventions of \cite{Nekrasov:2003rj}, we have
\be
a_{Dj}\,\equiv\,\frac{1}{2\pi i }\,\frac{\partial{\cal F}(\hat a)}{\partial\hat a_j}\,,\qquad j\,=\,1,2,\ldots N\,.
\ee
As defined previously the Coulomb branch moduli $\hat a_j \,=\,i a_j$ so that
\be
a_{D\,j}\left(ia,\,iM,\,i\tau\right)
\,+\, \overline {a_{D\,j}\left(ia,\,iM,\,i \tau\right)}\,=\,0\,.
\ee
With $a_j \in {\mathbb R}$, the saddle-point conditions are then concisely,
\be
{\rm Re}(a_{D\, j})\, =\, {\rm Re}(\hat a_j)\,=\,0\,,
\ee
for all $j$. This means that at putative saddle-points, the periods and dual periods must be `aligned' with the same complex phase and in particular, along the imaginary axis. More generally, when such an alignment of the phases of the periods occurs, one encounters a curve or wall of marginal stability along the Coulomb branch of ${\cal N}=2$ supersymmetric gauge theory \cite{sw}. Therefore the large volume 
saddle-points of the partition sum on $S^4$ can be viewed as special points on the curves of marginal stability where ${\rm Re}(a_j)=0$.

For special values of $\theta_{\rm YM}$ ($0$ or $\pi$), these may coincide with certain points of (maximal) degeneration of the Donagi-Witten curve where the periods are similarly aligned, leading to massless BPS dyons. Such points which descend to specific oblique confining vacua of ${\cal N}=1^*$ theory, can be described exactly for any $N$ and their contribition to the partition function can be computed exactly.

\subsection{Pure ${\cal N}=2$ SYM}
Before examining the ${\cal N}=2^*$ theory, we first focus attention on the simpler case of pure ${\cal N}=2$ SYM, which  is a special limit of ${\cal N}=2^*$ theory obtained by decoupling the adjoint hypermultiplet. For $SU(2)$,  ${\cal N}=2$ SYM the prepotential is 
\be
{\cal F}(\hat a)\,=\,-\,\frac{1}{2}\,\hat a^2\,\ln\left(\frac{\hat a}{\Lambda}\right)^2\,+\,{\cal F}_{\rm inst}(\hat a)\,,\qquad\Lambda \in {\mathbb R}\,.
\ee
We take the dynamical scale $\Lambda$ to be real, which is equivalent to setting the microscopic vacuum angle to zero.
The prepotential respects the symmetry under the Weyl group of $SU(2)$ which acts by permutation on the moduli $\hat a_{1,2}$ or equivalently as $\hat a\to - \hat a$. 
A branch cut singularity arises from the one-loop term, while the instanton contributions are even functions of $\hat a$, so that for large $\hat a$ we have
\be
{\cal F}_{\rm inst}(\hat a)\,=\,\hat a^2\sum_{k=1}^\infty\left(\frac{\Lambda}{\hat a}\right)^{4k}\,{\cal F}_k\,.
\ee
In terms of the microscopic parameters at the UV cutoff, $\Lambda^4\,=\,\Lambda_{\rm UV}^4\,\exp(-8\pi^2/g^2_{\rm YM})$. Pestun's formula for the partition function on $S^4$ instructs us to perform the integral along the imaginary axis in the complex $\hat a$-plane.  Taking $\hat a\,=\, i a$, we split the prepotential into its real and imaginary parts,
\bea
&&{\rm Re}\left[{\cal F}(ia)\right]\,=\,\frac{1}{2}\,a^2\ln\left(\frac{a^2}{\Lambda^2}\right)\,+\,{\cal F}_{\rm inst}(ia)\,,\\\nonumber\\\nonumber
&&{\rm Im}\left[{\cal F}(ia)\right]\,=\,\frac{\pi}{2}\,a^2\,.
\eea
The dual period $a_D$  defined as\footnote{The normalisations and conventions we use in this paper follow those adopted in \cite{Nekrasov:2003rj}. In particular $(2\pi i )a_{D\,j} \,=\,\partial{\cal F}/{\partial a_j}$, and for the $SU(2)$ theory $a_{D}\equiv a_{D1}-a_{D2}$.}
\be
a_{D}\,=\,\frac{1}{i\pi}\frac{\partial {\cal F}(\hat a)}{\partial \hat a}\,.\label{ad}
\ee
At the saddle point, the real part of $a_D$ is set to zero, and therefore we find
\be
a_D\,+\,\hat a\,=\,0\,,\qquad \qquad \hat a\,=\, ia\,.
\ee
This is the condition for degeneration of the Seiberg-Witten curve for $SU(2)$ and the appearance of a massless BPS dyon with magnetic and electric charges given as $(n_m, \, n_e)\,=\,(1,1)\,$. In particular, both $a_D$ and $\hat a$ are aligned along the imaginary axis and the point lies on the curve of marginal stability. This can be explicitly checked using the exact solution for $a_D$ and $\hat a$ in \cite{sw} which yields 
\be
a_D\,=\,-\hat a \,=\, -\frac{4i}{\pi}\Lambda\,.
\ee
The degeneration point where the $(1,0)$ BPS monopole becomes massless corresponds to $a_D=0$ and $\hat a\,=\, 4\Lambda/\pi \,\in {\mathbb R}$. This is a saddle-point of the integrand in \eqref{zlargevol}, when analytically continued away from the imaginary axis in the $\hat a$-plane. The dominant saddle-point is determined by the value of the real part of the prepotential. It can be readily verified that the critical point on the imaginary axis with a massless $(1,1)$ dyon, has lower action and is therefore dominant.

The analysis above generalises straightforwardly to the pure ${\cal N}=2$ theory with $SU(N)$ gauge group.
The prepotential for the pure $SU(N)$ theory is
\bea
&&{\cal F}(\hat a)\,=
-\frac{1}{2}\sum_{k< j}\hat a_{kj}^2\ln\left(\frac{\hat a_{kj}}{\Lambda}\right)^2\,+\,{\cal F}_{\rm inst}(\hat a)\,,\\\nonumber\\\nonumber
&&\hat a_{kj}\,=\,\hat a_k-\hat a_j\,,\qquad\qquad \Lambda \in {\mathbb R}\,.
\eea
Using the Weyl group of $SU(N)$, we can pick a specific ordering of the Coulomb branch moduli:
\be
a_1\,\geq\, a_2\,\geq\,\ldots\, a_{N-1}\,\geq\, a_N\,,\qquad \qquad \hat a_j\,=\,ia_j\,.
\ee
We then find 
\bea
&&a_{D\,j,\,j+1}\,+\,\frac{N}{2}\,\hat a_{j,\,j+1}\,=\,0\,,
\label{sp}\\\nonumber
&&a_{D\,j,\,j+1}\,\equiv\,a_{D\,j}\,-\,a_{D\,j+1}\,,
\qquad a_{D\,j}\,\equiv\,\frac{1}{2\pi i}\,\frac{\partial {\cal F}}{\partial \hat a_j}\,,
\eea
with $j=1,2\ldots N-1$. When $N$ is an even integer, these are precisely the conditions for the  appearance of $N-1$ massless BPS dyons. In particular the dyons each carry magnetic and electric  charge $\left(1,\frac{1}{2}N\right)$ under a distinct abelian factor on the Coulomb branch, and the Seiberg-Witten curve degenerates maximally at this point.  Note that the solution with $a_{Dj}=0$ is also a saddle point of the integrand analytically continued off the imaginary axis.

For $N$-odd and $\Lambda \in {\mathbb R}$, the conditions \eqref{sp} pick out a  specific point on the marginal stability curve which does not correspond to a singular point, although the ratios of the periods yield a rational number. This is because, in this case, there are no semiclassical bound states of dyons with magnetic charge 2 (see e.g. \cite{Fraser:1996pw}).
On the other hand if we introduce a microscopic (UV) theta-angle with $\theta_{\rm YM}=\pi$, we obtain
\be
\Lambda\, \to\, \Lambda\,e^{{i\pi}/{2N}}\,,\qquad  \Lambda \in {\mathbb R}\,,
\ee
and the saddle-point satisfies 
\be
a_{D\,j,\,j+1}\,+\,\frac{N-1}{2}\,\hat a_{j,\,j+1}\,=\,0\,, \qquad j\,=\,1,2,\ldots N-1\,.
\ee
Therefore when $N$ is an odd integer, these are the conditions for maximal degeneration  i.e. for $(N-1)$ massless BPS dyons,  each with charge $\left(1,\tfrac{N-1}{2}\right)$ under one distinct abelian factor on the Coulomb branch.

\subsection{Saddle-points for ${\cal N}=2^*$  theory}
\label{sec:saddles}

We now turn to the ${\cal N}=2^*$ theory. The physical picture of the saddle-points of the large volume partition function now has a new ingredient. Since the (complex) mass parameter for the adjoint hypermultiplet is imaginary, the point on the Coulomb branch where purely electric BPS states become light, occurs on the imaginary axis, i.e. whenever any of the differences $\hat a_{jk}$ is equal to $\pm i M$. Going around this point produces a monodromy which in turn implies that the physical interpretation of putative saddle-point configurations can depend on their location relative to this singularity.  

\subsubsection{The $SU(2)$ theory}
The $SU(2)$ theory happens  to exhibit some of the key features that generalise automatically and so we begin by focussing attention on this. Higher rank cases have a richer structure of putative saddle-points.   

The $SU(2)$  ${\cal N}=2^*$ theory has  3 singularities on the Coulomb branch \cite{sw2, dw}. As is well known, when the theory is deformed by an ${\cal N}=1$ SUSY preserving mass term for the chiral multiplet residing in the ${\cal N}=2$ vector multipet, these three points descend to the three massive vacua of ${\cal N}=1^*$ theory with $SU(2)$ gauge group. The vacua realise three distinct phases, namely, Higgs $(H)$, confinement $(C)$ and oblique confinement $(C^\prime)$, corresponding to the condensation of the $(0,1)$ adjoint hypermultiplet, $(1,0)$ BPS monopole and $(1,1)$ BPS dyon respectively.   The $SL(2,{\mathbb Z})$ duality of ${\cal N}=4$ theory permutes the three phases.

We will denote the locations of these three points on the Coulomb branch in terms of the gauge-invariant coordinate
\be
u_2\,=\,\langle{\rm Tr}\Phi^2\rangle\,,
\ee
as $u_H$, $u_C$ and $u_{C^\prime}$. The prepotential for the theory has the form
\bea
{\cal F}(\hat a)\,=&&-\frac{1}{2}\left[\hat a^2\ln \hat a^2\,-\,\tfrac{1}{2}
\left(\hat a-iM\right)^2\ln\left(\hat a-iM\right)^2\,-\,\tfrac{1}{2}\left(\hat a+iM\right)^2\ln\left(\hat a +iM\right)^2\right]\nonumber\\\nonumber\\
&&+\,\frac{i\pi\tau}{2}\, \hat a^2\,+\,{\cal F}_{\rm inst}(\hat a)\,,\qquad\qquad \hat a\,\equiv\,\hat a_1-\hat a_2\,,\qquad
{\theta}_{\rm YM}\,=0\,.
\eea
\begin{figure}
\begin{center}
\includegraphics[width=2.75in]{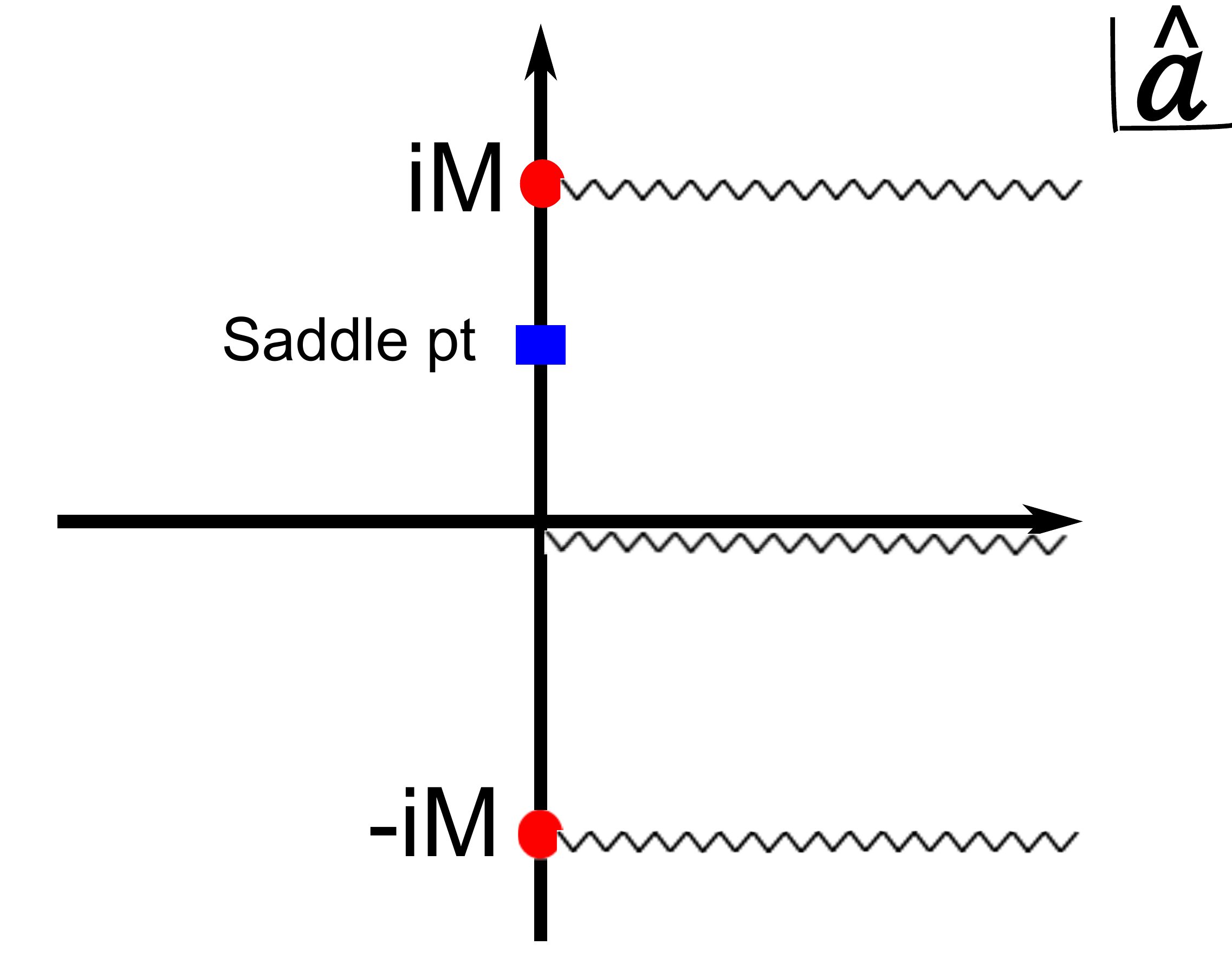}
\end{center}
\caption{\small { Complex $\hat a$-plane with the contour integral along the imaginary axis $\hat a\, =\, i a$. Also depicted are the branch cuts singularities of the prepotential ${\cal F}(ia)$.}}
\label{aplane}
\end{figure}
Along the imaginary slice $\hat a\,=\, i a$, the prepotential has both imaginary and real parts.
While the real part of ${\cal F}(ia)$ is obtained  by taking a principal value, the imaginary part is a discontinuous function of $a$. With $\theta_{\rm YM}=0$, the instanton action $q=\exp(2\pi i\tau)$ is real and since ${\cal F}_{\rm inst}(ia)$ respects the Weyl reflection symmetry, it is 
a function of $\hat a^2$ and is also real (see e.g.\cite{billo}). The imaginary part of the prepotential depends on the choice of orientation of the branch cuts of the one-loop contributions. The orientation of branch cuts must respect the requirement that for large $\hat a$, the theory  reduces to ${\cal N}=4$ SYM:
\be
{\cal F}(\hat a)\quad\to\quad \frac{i\pi \tau}{2}\,\hat a^2\,,\qquad\qquad{|\hat a|\gg |M|}\,.
\ee
Since the Weyl symmetry identifies the points $\hat a$ and $-\hat a$, without loss of generality, we take $\hat a =i a$ with $a>0$ and $M>0$.  With the branch cuts of ${\cal F}(ia)$ chosen as in fig.\eqref{aplane}, we then have (for $\rm \theta_{\rm YM}=0$):
\begin{itemize}
\item{ {$\bf a\,>\,M$:}
\bea
{\rm Re}\left[{\cal F}(ia)\right]\,=&& {F}_{>}(a)\\\nonumber\\\nonumber
=&&\frac{1}{2}\left[a^2\ln a^2\,-\,\tfrac{1}{2}(a+M)^2\ln(a+M)^2\,-\,
\tfrac{1}{2}(a-M)^2\ln(a-M)^2\right]
\\\nonumber
&&+\,\tfrac{2\pi^2}{g^2_{\rm YM}}\,a^2\,+\,{\cal F}_{\rm inst}(ia)\,.\\\nonumber
\\\nonumber
{\rm Im}\left[{\cal F}(ia)\right]\,= && 0\,.
\eea
}
\item{ {$\bf 0\,<\,a\,<\,M$:}
\bea
{\rm Re}\left[{\cal F}(ia)\right]\,=&& F_{<}(a)\\\nonumber\\\nonumber
=&&\frac{1}{2}\left[a^2\ln a^2\,-\,\tfrac{1}{2}(M+a)^2\ln(M+a)^2\,-\,
\tfrac{1}{2}(M-a)^2\ln(M-a)^2\right]\nonumber
\\
&&+\,\tfrac{2\pi^2}{g^2_{\rm YM}}\,a^2\,+\,{\cal F}_{\rm inst}(ia)\,.\\\nonumber
\\\nonumber
{\rm Im}\left[{\cal F}(ia)\right]\,
 = &&\frac{\pi}{2}\left(-a^2\,+\,2aM\,-\,2M^2\right)\,.
\eea
}
\end{itemize}
The Pestun partition sum is determined by the minimum of  ${\rm Re}\left[{\cal F}(ia)\right]$, the real part of the prepotential. However, the physical interpretation of the extremal point becomes apparent upon examination of the full holomorphic function, evaluated on the imaginary axis. In particular, the interpretation of  critical points will depend on their location relative to the singular point $H$ where $a=M$ and where the adjoint hypermultiplet becomes massless\footnote{It is also possible for critical points to lie in the complex plane and their contributions can be picked up by deforming the original integration contour smoothly.}.

\paragraph{Critical point for $a<M$:}  
This region is connected to the pure ${\cal N}=2$ theory in the decoupling limit $M\to \infty$ and $g^2_{\rm YM}\to 0$, whilst keeping fixed   $\Lambda \sim M\exp(-2\pi^2/g^2_{\rm YM})$. We define $a_D$ as 
\be
a_D\,=\,\frac{1}{i\pi}\,\frac{\partial {\cal F}(\hat a)}{\partial \hat a}\,+\,iM\,.\label{newdef}
\ee
The shift by $iM$, which is confusing at first sight, 
can be attributed to the monodromy around $a=iM$, which leads to a shift ambiguity (linear in $M$) in the period integral of the Seiberg-Witten differential \cite{sw2, Ferrari:1996de, Konishi:1998mk}. With this definition, it is easy to check that  the (1,0) monopole singularity in the decoupling limit, appears at $\hat a \sim \Lambda \in {\mathbb R}$ and corresponds to the condition $a_D\,=\,0$, as expected in the pure ${\cal N}=2$ theory. 

\begin{figure}
\begin{center}
\includegraphics[width=2.75in]{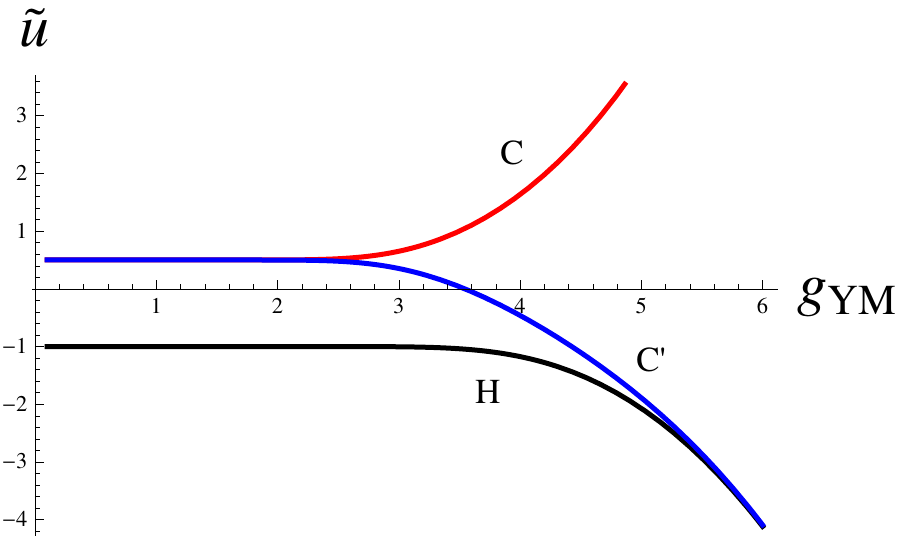}
\end{center}
\caption{\small {Positions of the coordinates $\tilde u$ of the singularities 
$C$ (red), $C^\prime$ (blue) and $H$ (black) as a function of the microscopic coupling $g_{\rm YM}$. Crucially the saddle-point $C^\prime$ (the dyon singularity) never crosses the hypermultiplet point $H$, where $\hat a\,=\,iM $.}}
\label{uvsg}
\end{figure}

The saddle-point condition becomes
\be
a_D(\hat a)\,-\,\hat a
\,=\,0\,,\qquad \hat a\,=\,ia\,,\qquad 0< a < M\,.
\ee
The resulting equation, in the decoupling limit, yields $\hat a \sim 
i\Lambda$, which is the singular point in ${\cal N}=2$ SYM where the $(1,1)$ BPS dyon becomes massless. This physical picture also holds away from the decoupling 
limit, as we will show in explicit detail in 
section \ref{sec:nekrasov}. 
The exact location of the dyon singularity $C'$ can be determined directly from the Seiberg-Witten curve \cite{sw2}:
\be
y^2\,=\,\prod_{i=1}^3\left(x\,-\,e_i(\tau)\, \tilde u\,+\,\frac{M^2}{4}e_i(\tau)^2 \right)\,,
\ee
where 
\be
\tilde u\,=\,\langle {\rm Tr}\Phi^2\rangle \,+\,\frac{M^2}{12}\left(1\,+\,\sum_{n=1}^\infty \alpha_n\,q^n\right)\,.
\ee
The $\{\alpha_n\}$ represent scheme-dependent, but vacuum-independent additive ambiguities \cite{dkm}. The locations of the three singular points are then given by,
\bea
&&\tilde u_H\,=\,-\frac{M^2}{4}e_1(\tau)\,=\,\frac{M^2}{6}\,\left[E_2(\tau)\,-\,2\,E_2(2\tau)\right]\,,\label{u2su2}\\\nonumber\\\nonumber
&&\tilde u_C\,=\,-\frac{M^2}{4}e_2(\tau)\,=\,\frac{M^2}{6}\,\left[E_2(\tau)\,-\,\tfrac{1}{2}\,E_2\left(\tfrac{\tau}{2}\right)\right]\,,\\\nonumber\\\nonumber
&&\tilde u_{C^\prime}\,=\,-\frac{M^2}{4}e_3(\tau)\,=\,\frac{M^2}{6}\,\left[E_2(\tau)\,-\,\tfrac{1}{2}\,E_2\left(\tfrac{\tau}{2}+\tfrac{1}{2}\right)\right]\,.
\eea
Here $\tau\,=\, 4\pi i/g^2_{\rm YM}$ and $E_2$ is the second Eisenstein series which is an ``almost'' modular form of weight two (see appendix \ref{app:elliptic} for details). Whilst the actual values of the coordinates are ambiguous, their relative locations are completely unambiguous (and real for $\theta_{\rm YM}=0$). At weak coupling $g_{\rm YM}\ll 1$, using the $q$-expansions \eqref{qexpansion}
\bea
 &&u_H\,-\,u_{C}\,\simeq -\frac{M^2}{4}\,-\,2M^2\,e^{-4\pi^2/g^2_{\rm YM}}\,,\\\nonumber\\\nonumber 
 && u_H\,-\,u_{C^\prime}\,\simeq -\frac{M^2}{4}\,+\,2M^2\,e^{-4\pi^2/g^2_{\rm YM}}\,,
\eea 
as expected from the results for the pure ${\cal N}=2$ theory. At strong coupling $g_{\rm YM}\gg 1$, we can apply the (anomalous) modular transformation rule for $E_2$ and obtain
\bea
&&u_H\,-\,u_{C}\,\simeq -\frac{M^2}{4}\,\left(\frac{g^2_{\rm YM}}{4\pi}\right)^2\,,\qquad g_{\rm YM}\gg 1\,,\\\nonumber\\\nonumber 
 && u_H\,-\,u_{C^\prime}\,\simeq -\,{4M^2}\,\left(\frac{g^2_{\rm YM}}{4\pi}\right)^2\,e^{-g^2_{\rm YM}/4}\,.
\eea
Therefore, both at weak and strong gauge coupling, the monopole and dyon singularities $C$ and $C'$ remain to one side of the point $H$ where the adjoint hypermultiplet is massless. The positions of the singularities are shown in fig.\eqref{uvsg}.
The main point of this exercise was to show that the saddle-point $C^\prime$ can never collide with $H$. The fact that maximally singular points on the ${\cal N}=2^*$ Coulomb branch (or massive vacua of ${\cal N}=1^*$ theory) cannot merge, was pointed out in \cite{dw}. This point has also been made by Russo \cite{russo} recently within the present context.

Therefore we conclude that there is one saddle-point $C^\prime$ on the axis ${\rm Re}(\hat a)=0$  with $a<M$, which exists for all values of $g_{\rm YM}$, and which descends  to the oblique confining vacuum of ${\cal N}=1^*$ theory. We will calculate the free energy of this saddle point using Nekrasov's functional in section \ref{sec:nekrasov}.

\paragraph{(No) critical point for $a>M$:}  The large-$a$ regime is smoothly connected to the semiclassical region where quantum corrections and instantons can be made small for sufficiently large $a$, and the theory approaches ${\cal N}=4$ SYM. We have already seen that the singular points $C$ and $C^\prime$ which lie on the real axis in the $u$-plane, never cross the  hypermultiplet point $(H)$ where $a=M$. Therefore a critical point, if any, in the large-$a$ regime cannot be a singular point. It is instructive to examine the prepotential to understand the conditions under which a critical point may exist for large $a$. With $a>M$, the one-loop prepotential is manifestly real. Using the definition of the dual period \eqref{newdef} which is compatible with the charges of light states at the singularities, the critical point condition for $a>M$ becomes
\be
a_D (\hat a)\,=\, iM\,, \qquad \hat a\,=\,ia\,.
\ee
Since this cannot be a singular point, it can only correspond to a point of marginal stability where ${\rm Im}(a_D/\hat a)\,=\,0$.

 Splitting ${\cal F}(ia)$ into the one-loop (including the classical piece) and instanton contributions,
\be
{\cal F}(ia)\,=\,{\cal F}_{\rm 1-loop}(ia)\,+\,{\cal F}_{\rm inst}(ia)\,,
\ee
it is easily seen that ${\cal F}_{\rm 1-loop}$ has a critical point at strong coupling. This occurs when the first derivative of ${\cal F}_{\rm 1-loop}$ becomes negative i.e. $g^2_{\rm YM}> 2\pi^2/\ln2 \simeq  28.48$:
\bea
&&{\cal F}_{\rm 1-loop}(ia)\,=
\\\nonumber
&&\qquad\tfrac{1}{2}\left(a^2\ln a^2\,-\,\frac{1}{2}(a+M)^2\ln(a+M)^2\,-\,\tfrac{1}{2}(a-M)^2\ln(a-M)^2\right)\,+\,\tfrac{2\pi^2}{g^2_{\rm YM}}a^2\,,
\eea
and
\be
{\cal F}_{\rm 1-loop}(ia)\,\simeq\,M^2\left[\tfrac{2\pi^2}{g^2_{\rm YM}}-\ln\left(4M\right)\right]\,+\,2M(a-M)\,\left(\tfrac{2\pi^2}{g^2_{\rm YM}}-\ln 2\right)+\ldots
\ee
If one-loop effects were dominant then this would lead to a minimum for $a>M$, since ${\cal F}(ia)$ must eventually turn around and increase as $a^2$ for large enough $a$.
 However, the instanton contributions are equally important for this value of the coupling. In particular, the form of the instanton prepotential is known \cite{Minahan:1997if} in the regime $a > M$:
\be
{\cal F}_{\rm inst}(ia)\,=\,\sum_{n=1}^\infty(-1)^{n+1}\frac{f_{2n}(\tau)}{(2n)}\,\frac{M^{2n+2}}{a^{2n}}\,.\label{largea}
\ee
The functions $f_{2n}(\tau)$ are given in terms of anomalous modular forms of weight $2n$. For example, $f_2(\tau)=(E_2(\tau)-1)/6$ and $f_4(\tau)\,=\,E_2^2/18\,+\,E_4/90\,-\,1/15$. In the weak coupling limit, the instanton prepotential vanishes, $f_{2n}\to 0$. At strong coupling, after applying an $S$-duality,
\be
f_{2n}(\tau)\,\sim\, \left(g_{\rm YM}\right)^{4n}\,,\qquad g_{\rm YM}\gg 1\,.
\ee
Therefore, at strong coupling, instanton terms (after S-duality) remain small only if
\be
a\,\gg g^2_{\rm YM}\,M\,.
\ee
Hence, we cannot use \eqref{largea} to conclude whether or not the critical point of the one-loop prepotential is washed out by the instanton part of the effective action. Interestingly, at arbitrarily strong coupling, ${\cal F}_{\rm 1-loop}$ continues to have a critical point:
\be
\left.\frac{\partial {\cal F}_{\rm 1-loop}(ia)}{\partial a}\right|_{a\gg M;\,g_{\rm YM \gg 1}}\,
=\,0\quad\implies\quad a\simeq \frac{g_{\rm YM}}{2\pi}\,M\,.
\ee
This is, however, deep within the region where ${\cal F}_{\rm inst}$ cannot be neglected (at strong coupling).

\begin{figure}
\begin{center}
\includegraphics[width=2.75in]{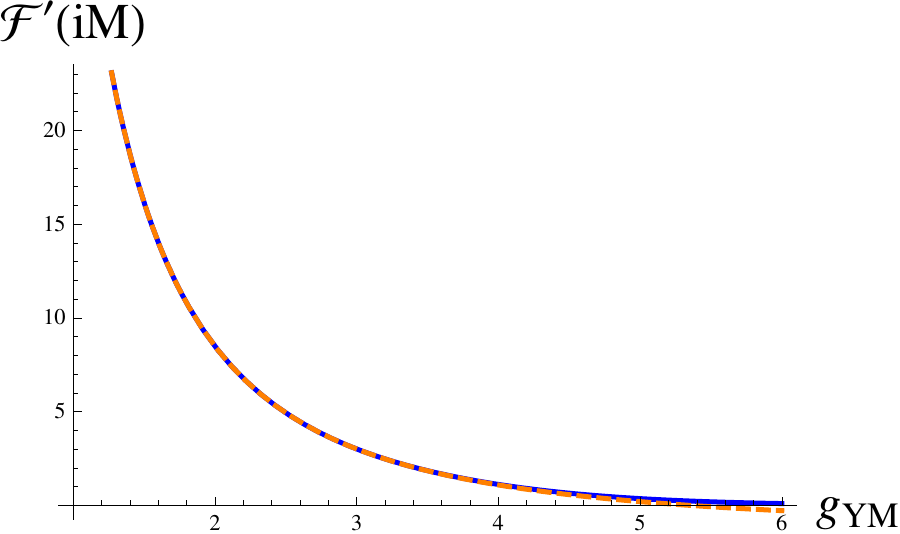}\hspace{0.2in}
\includegraphics[width=2.75in]{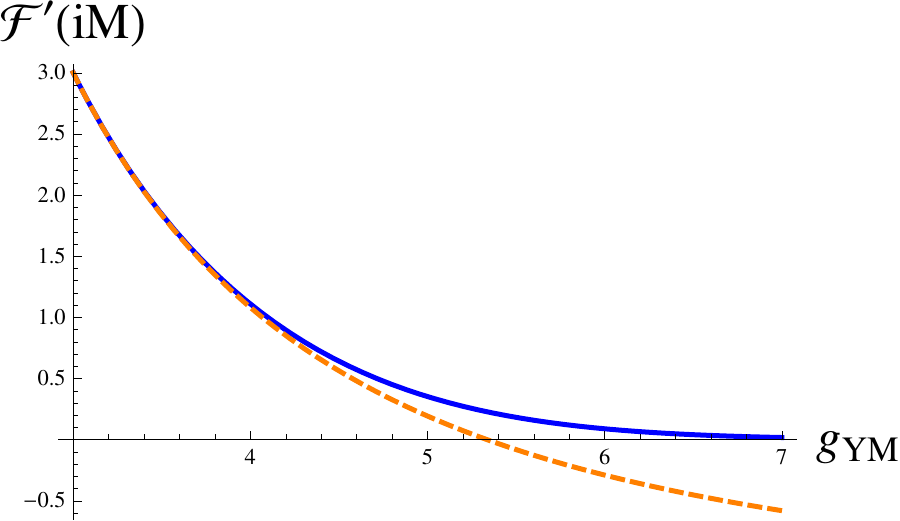}
\end{center}
\caption{\small {The slope of the one-loop prepotential (dashed, orange) and the full prepotential (solid, blue) at $a=M$, as a function of the gauge coupling $g_{\rm YM}$. The two curves are practically indistinguishable (left) until the instanton contributions kick in (right) and prevent  ${\cal F}'(iM)$ from becoming negative for any value of $g_{\rm YM}$}}
\label{slope}
\end{figure}

To determine whether the critical point of the one-loop prepotential survives the inclusion of instanton corrections, we need to know the instanton expansion about the singular point $a=M$. Such an expansion was considered by Minahan et al in \cite{Minahan:1997if} and the leading term in ${\cal F}''$ was identified exactly. 
 We first define a formal expansion of ${\cal F}_{\rm inst}$ around the singular point, in powers of $(a^2-M^2)$:
\bea
&&{\cal F}_{\rm inst}(ia)\,=\,M^2\,c_0(q)\,+\,(a^2 -M^2)\,c_1(q)\,+\,\frac{1}{M^2}(a^2-M^2)^2\, c_2(q)\,+\ldots\\\nonumber
&&q\,=\,e^{2\pi i\tau}\,.
\eea
The constant term $c_0(q)$ is irrelevant for our purpose. Using the results in \cite{Minahan:1997if} for the explicit form of the large-$a$  expansion \eqref{largea}, the instanton expansion to order $q^8$, and the exact formula for ${\cal F}''(iM)$, we deduce that
\bea
&& c_1(q)\,=\,-2\ln\prod_{n=1}(1+q^n)\,-\,4\ln\prod_{n=1}\left(1+(-q)^n\right)\\\nonumber\\\nonumber
&&c_2(q)\,=\,-\ln\prod_{n=1}\left(\frac{1+q^n}{1+(-q)^n}\right)\,.
\eea
Therefore, near the hypermultiplet point, combining classical, one-loop and all instanton corrections we obtain,
\be
{\cal F}(ia) -{\cal F}(iM)\,\approx\,M(a-M)\left[- 2\ln 2\,-\,4\ln\frac{\eta(2\tau)}{\eta(\tau)}\,-\,8\ln\frac{\eta(2\tau)}{|\eta(\tau+\frac{1}{2})|}\right]+\ldots 
\ee
where $\eta(\tau)\,=\,e^{i\pi\tau/12}\prod(1-q^n)$ is the Dedekind eta-function. Although not of immediate relevance, we note in passing that ${\cal F}(iM)$ can be written in closed form as
 \be
 {\cal F}(iM)\,=\,2M^2\left[\ln\eta(\tau)\,-\,2\ln\eta(2\tau)\,-\,\ln(2M)\right]\,.
 \ee
 It can now be seen explicitly that whilst ${\cal F}_{\rm 1-loop}'(iM)\,=(\,-2\ln2-i\pi\tau)M$ becomes negative for $g_{\rm YM}\gtrsim 5.34 $, the inclusion of all instanton corrections forces ${\cal F}'(iM)$ to be strictly greater than zero (see fig.\eqref{slope}). Although this does not exclude the possibility of a critical point for $a $ significantly larger than $M$, it appears quite unlikely. 

We have argued that the partition function of the ${\cal N}=2^*$ theory with $SU(2)$ gauge group, on a large four-sphere, is computed by a single saddle point (the dyon singularity $C^\prime$) and therefore the system cannot exhibit any non-analyticities as a function of the gauge coupling. This was also the expectation in \cite{russo}. 
The value of the partition function at this saddle point will be evaluated using Nekrasov's functional. 

\subsubsection{$SU(N)$ ${\cal N}=2^*$ theory with $N>2$}
\label{sec:N=2*}

For $SU(N)$ gauge group, with $N>2$, and $N$ an {\em even } integer, we find new putative saddle point configurations, in addition to generalisations of the oblique confining and confining points that appeared for $SU(2)$. The prepotential for ${\cal N}=2^*$ theory is
\bea
{\cal F}(\hat a)\,=
-\frac{1}{2}\sum_{k< j}&&\left[\hat a_{kj}^2\ln\hat a_{kj}^2\,-\,\tfrac{1}{2}
\left(\hat a_{kj}+iM\right)^2\ln \left(\hat a_{kj}+iM\right)^2\,\right.\\\nonumber
&& \left.-\,\tfrac{1}{2}\left(\hat a_{kj}-iM\right)^2\ln \left(\hat a_{kj}-iM\right)^2
\right] -\,\tfrac{4\pi^2}{g^2_{\rm YM}}\sum_j \hat a_j^2\,+\,{\cal F}_{\rm inst}(\hat a)\,,\\\nonumber
&&\hat a_{kj}\,=\,\hat a_k-\hat a_j\,,\qquad \theta_{\rm YM}=0\,.
\eea
Focussing attention on the imaginary $\hat a$-axis and choosing a natural ordering for the $a_j$ as explained for the pure ${\cal N}=2$ theory,
we find that putative critical points can be summarized as follows:
\begin{itemize}
\item{For small enough, real $a_{ij}$ such that $a_{ij}<M$ for all $i,j$, we find that the saddle-point conditions imply,
\bea
a_{D\,j,\,j+1}\,=\,\frac{N}{2}\,\hat a_{j,\,j+1}\,,
\label{sp1}
\eea
with $j=1,2\ldots N-1$. As in the $N=2$ case, we have absorbed a linear shift $i M$ into the definition of the dual periods $a_{Dj}$. For  $N$ even, we recognise these as the conditions for the  appearance of $N-1$ massless BPS dyons, each carrying charges $(n_m^{(j)}, n_e^{(j)})\,=\,\left(1, - \frac{N}{2}\right)$, $(j=1,2\ldots N-1)$, in a distinct unbroken $U(1)$ subgroup on  the Coulomb branch.  This is smoothly related to the oblique confining point we saw above for the pure ${\cal N}=2$ theory.
}
\item{If all $a_{ij}$ are large such that $a_{ij}> M$, then saddle-point conditions pick out the point satisfying
\be
a_{D\,j,\,j+1}\,=\,iM\,,\qquad\qquad j=1,2\ldots N-1\,.
\ee
We have already seen in the $SU(2)$ theory that such a saddle point is unlikely to exist. 
}
\item{Finally, there is potentially a large family of critical points where a subset of $a_{ij}$ are smaller than, and the rest are larger than $M$. The simplest of these situations arises when $a_{1N} > M$ and all other $a_{ij}<M$. A putative saddle point with this property would need to satisfy
\bea
&& a_{D\,12}\,=\,\frac{N}{2}\hat a_{12}\,-\,\frac{1}{2}\left(\hat a_{\,1N}-iM\right)\,,\label{mixed}
\\\nonumber\\\nonumber
&& a_{D\,j,\,j+1}\,=\,\frac{N}{2}\hat a_{j,\,j+1}\,,\qquad j\,=\,2,3,\ldots N-2\,.
\\\nonumber\\\nonumber
&& a_{D\,N-1,\,N}\,=\,\frac{N}{2}\hat a_{N-1,\,N}\,-\,\frac{1}{2}\left(\hat a_{\,1N}-iM\right)\,.
\eea
These are no longer conditions  for maximal degeneration.  A subset of the dual periods are degenerate and lead to massless dyons, but $a_{D\,12}$ and $a_{D\,N,N-1}$ are required to be (non-integer) linear combinations of cycles with non-zero intersection. 
This picks out a particular point on the wall/surface of marginal stability in the ${\cal N}=2^*$ Coulomb branch. 

We have only considered the simplest such `mixed' saddle-point. It should be fairly clear that there is a large family of such possible saddle-points with increasing $N$.
Whether there exist points on the Coulomb branch which actually satisfy these conditions is a dynamical question that will require analysis on a case-by-case basis, and will be a function of $N$ and the gauge coupling $g_{\rm YM}$, as already illustrated for the $SU(2)$ theory. One may generically expect at least some of these saddle points to co-exist, leading to phase transitions as a function of $g_{\rm YM}$. This is consistent with the results of \cite{zaremborusso1, zaremborusso2} where the large-$N$ limit was analysed and the theory argued to exhibit an infinite sequence of phase transitions as a function of increasing 't Hooft coupling.

}
\end{itemize}

Similarly to the pure ${\cal N}=2$ case, the critical points are related to singular points only for special values of $\theta_{\rm YM}$. When $N$ is odd, and $\theta_{\rm YM}\,=\pi$, the large-$a$ and small-$a$ critical-point conditions become
\bea
&& a_{D j,\,j+1}\,=\,\frac{N+1}{2}\hat a_{j,\,j+1}\,,\qquad a_{j,j+1}< M \label{Noddsmall}\\\nonumber\\
&&a_{D j,\,j+1}\,=\,iM\,+\,\hat a_{j,\,j+1}\,,\qquad a_{j,j+1}>M\,.\label{Noddlarge}
\eea
with $j=1,2,\ldots N-1$ and $\hat a_j\,=\,i\,a_j$. In addition to these, there is potentially a large family of `mixed' critical points with a certain number of periods $a_{j,\,j+1}$ small and the rest being large. The condition \eqref{Noddsmall} implies maximal degeneration of the Donagi-Witten curve and the appearance of massless BPS $(1,\tfrac{1}{2}(N+1))$ dyons in each abelian factor.

\section{The Nekrasov partition function and critical points}
\label{sec:nekrasov}
It turns out that the contributions of the maximally degenerate saddle-points to the partition function can be computed exactly for any $N$ and any value of the microscopic gauge coupling $g^2_{\rm YM}$. 
 For this purpose, the most significant aspect of the Nekrasov partition function in the limit 
$\epsilon_{1,2}=R^{-1}\to 0$ is that it is  dominated by a saddle-point of the functional \cite{Nekrasov:2003rj, Hollowood:2003cv}
\bea
{\cal E}_\tau[\rho,\lambda,\hat a,\,iM]&&=\,\nonumber\\\nonumber
-N^2&&\int_{\cal C}dx \,dy\,\rho(x)\left[\gamma_0(x-y)-\tfrac{1}{2}\gamma_0(x-y-iM)-\tfrac{1}{2}\gamma_0(x-y+iM)\right]\rho(y)\nonumber\\
 &&+\,i\pi\tau N\int_{\cal C} dx\, x^2\,\rho(x)\,+\,
\sum_{j=1}^N\lambda_j\left(\hat a_j\,-\,N\int_{{\cal C}_j}dx\, x\,\rho(x)\right)\,,
\label{action}\\\nonumber\\
\gamma_0(x)\,&&=\,\tfrac{1}{4}\,x^2 \ln x^2\,,\\\nonumber\\
{\cal C}\,&&\equiv\,\bigcup_{j=1}^N\, {\cal C}_j\,,\qquad  {\cal C}_j\,=\,[\alpha_j^-, \alpha_j^+]\,.
\eea
The function $\rho(x)$ is a density with support on the disjoint union of $N$ intervals $\{{\cal C}_j\}$, satisfying
\be
\int_{{\cal C}_j}dx\,\rho(x)\,=\,\frac{1}{N}\quad \forall j\,,\qquad \int_{{\cal C}}dx\,\rho(x)\,=\,1\,,\label{norm}
\ee
whilst the $\{\lambda _j\}$ are
$N$ Lagrange multipliers enforcing the constraints
\be
\hat a_j\,=\,N\int_{{\cal C}_j}dx\, x\,\rho(x)\,,\qquad{j=1,2,\dots N}\,.
\ee
For a fixed set $\{\hat a_j\}$, specifying a Coulomb branch configuration, the instanton partition sum is simply
\be
{\cal Z}_{\rm Nekrasov}\,\sim\,\exp\left(-R^2\,{\cal E}_\tau[\rho]\right)\,.
\ee
In the language of \cite{Nekrasov:2003rj}, the instanton partition function for $\epsilon_{1,2}\neq 0$, can be written as a sum over coloured partitions, to each of which is associated a piecewise-linear ``path'' $f(x)$. In the limit $\epsilon_{1,2}\to 0$, the path $f(x)$ becomes smooth, and the sum over paths localizes onto saddle-points of the above functional with the density function related to $f(x)$ as
\be
\rho(x)\,=\,\frac{1}{2N}\,f''(x)\,.
\ee
\begin{figure}
\begin{center}
\includegraphics[height=1.39in]{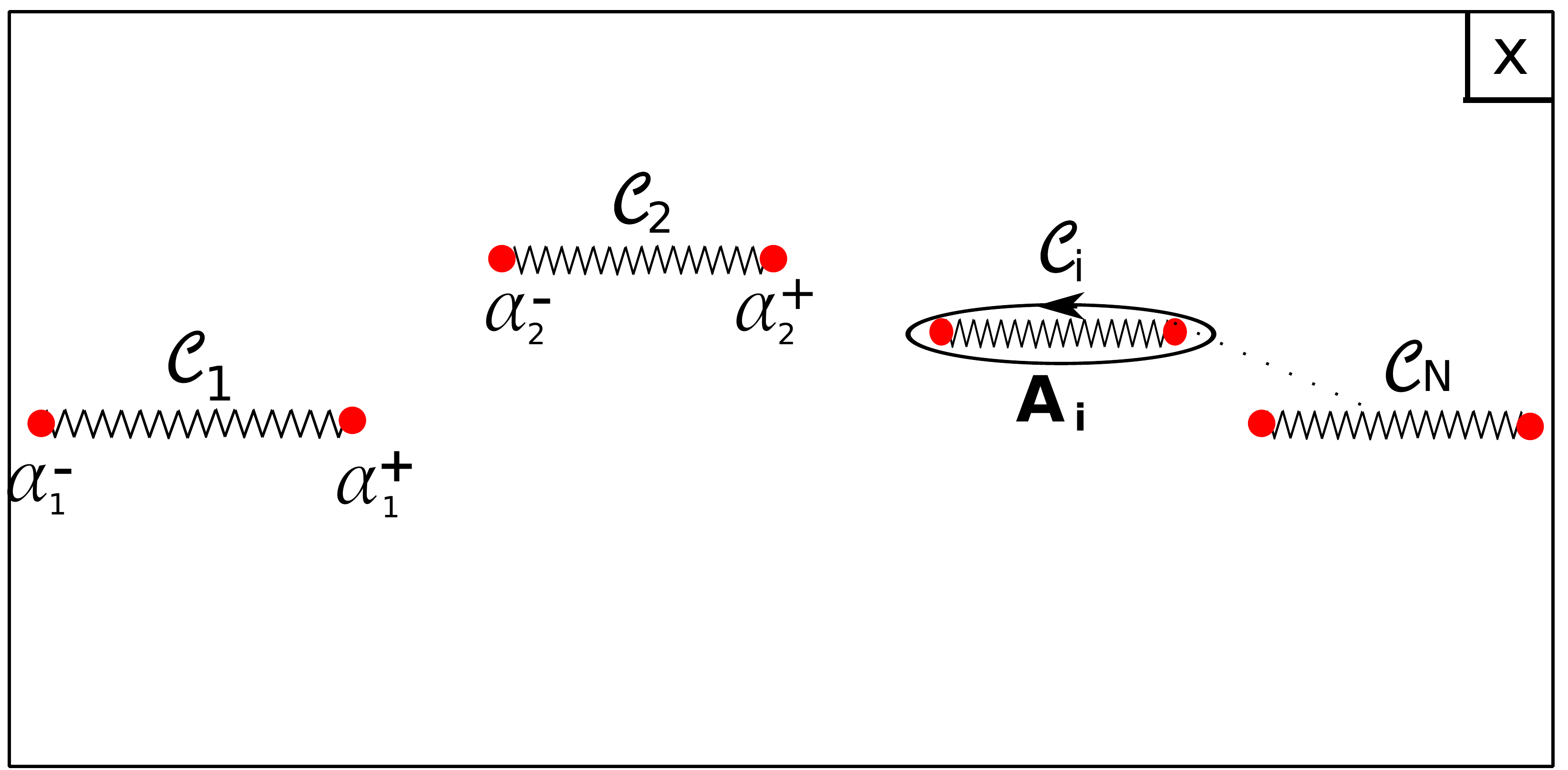}
\hspace{0.1in}
\includegraphics[width=2.9in]{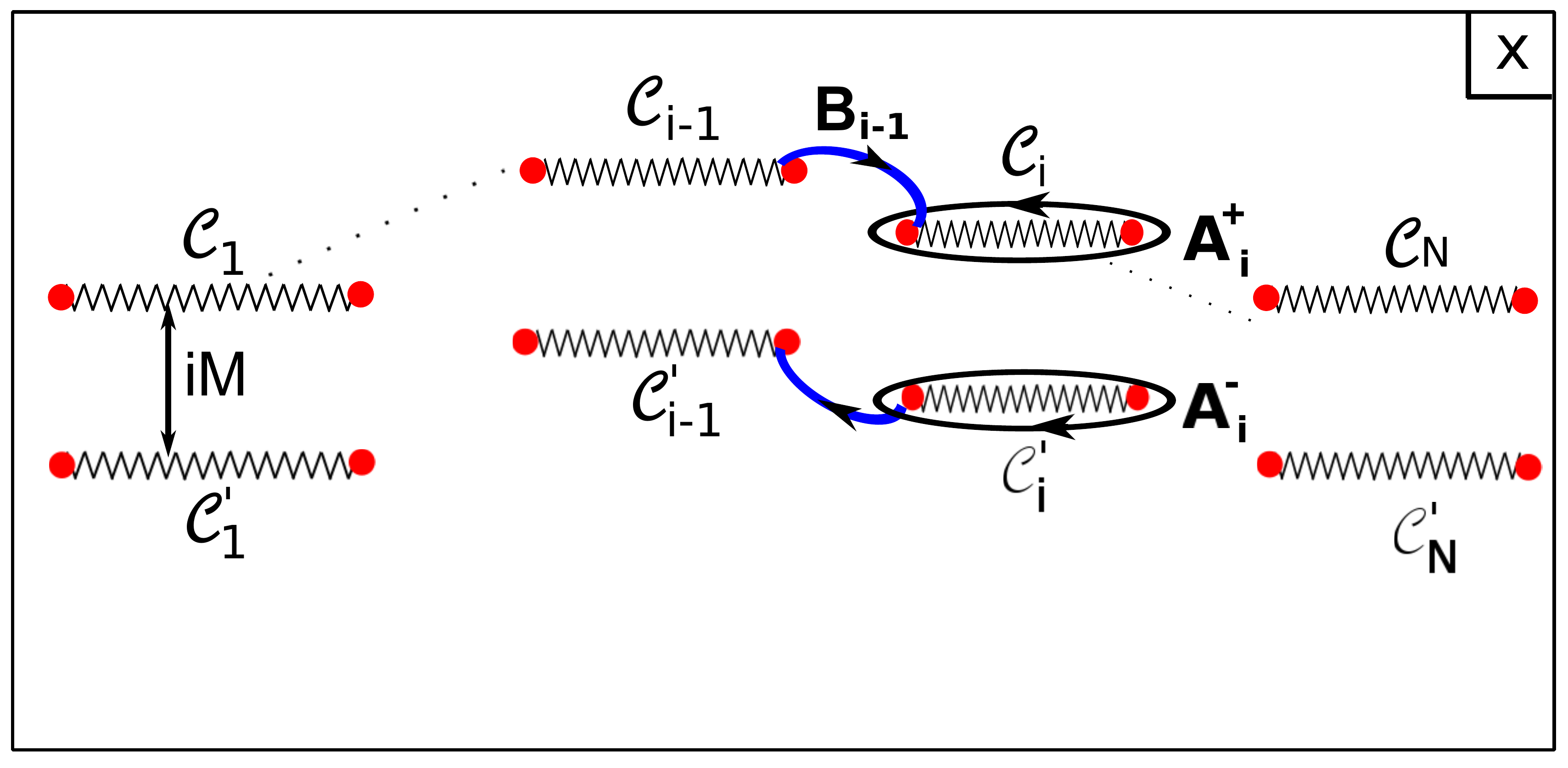}
\end{center}
\caption{\small {{\bf Left:} The $N$ intervals $\{{\cal C}_i\}$ in the complex $x$-plane where $\rho(x)$ has support. {\bf Right:} The genus $N$ Riemann surface associated to the function $G(x)$, with gluing conditions for each of the $N$ branch cuts $\{{\cal C}_i\}$ with their respective images $\{{\cal C'}_i\}$ (shifted by $-i M$). Points just above (below) ${\cal C}_i$ are identified with corresponding points just below (above) ${\cal C}_i'$.
}}
\label{gfig}
\end{figure}
Specifically, the function $f(x)$ determines the limit shape of a Young tableaux which characterises the representation dominating the instanton partition sum in the limit $\epsilon_{1,2}\to 0$, when the number of boxes in the tableaux diverges.

We note that the kernel appearing in the action functional \eqref{action} 
is precisely the  one-loop prepotential for ${\cal N}=2^*$ theory. 
At fixed $N$, the localisation to saddle-points of the functional ${\cal E}_\tau[\rho]$ is achieved by the large volume or small $\epsilon_{1,2}$ limit.

The partition function of  the theory on $S^4$ involves integration over the $\{\hat a_j\}$, in addition to the integrations over the Lagrange multipliers. Since the exponent of the instanton partition function  scales as $R^2$, the measure factor in \eqref{zs4} is subleading  in the large volume limit, and the partition function can be evaluated on the saddle-point(s) of the integrand of
\be
{\cal Z}_{S^4} \sim \int [da]  [d\lambda] [d\tilde\lambda]
 [d\rho] [d\tilde \rho]\,\exp\left[-R^2\left\{{\cal E}_\tau[\rho,\,ia,\,\lambda, iM ]+{\cal E}_{\bar \tau}[\tilde \rho,\,- ia,\,\tilde\lambda, - iM ]\right\}\right]\,.\label{largevol}
\ee
Pestun's matrix integral \eqref{zs4} involves two copies of the Nekrasov instanton partition sum. Thus we have two energy functionals to extremize in the large volume limit with {\it a priori} independent density functions $\rho, \tilde\rho$ and Lagrange multipliers $\lambda, \tilde\lambda$.  An important feature of the partition sum is that the moduli $\hat a_j$, must be taken to be purely imaginary,
\be
\hat a_j\,=\, i a_j\,.
\ee
The extremization conditions will then relate the moduli and, therefore, the density functions of the two copies.
Varying independently with respect to each set of variables we obtain the following set of saddle-point equations
\bea
&& \lambda_j\,=\,\tilde \lambda_j\,,\hspace{3in} j=1,2,\ldots N\,,\label{lambdaeq}
\\\nonumber\\
&&ia_j\,=\, N\int_{{\cal C}_j}dx\, x\, \rho(x)\,=\,- N\int_{\tilde{\cal C}_j}
d\tilde x\, \tilde x\,\tilde\rho(\tilde x)\,,\label{aeq} \\\nonumber\\\nonumber
&& \frac{\lambda_j}{N}\,=\,-\int_{\cal C} dy\,\left[K(x-y)\,-\,\tfrac{1}{2}K\left(x-y- iM\right)\,-\,\tfrac{1}{2}K\left(x-y+iM\right)\right]\rho(y)+ \tfrac{2i\pi\tau}{N}\,x\,,\\\nonumber\\
&&\hspace{4.6in} x\in {\cal C}_j\,,\label{adeq}\\\nonumber\\\nonumber
&&\frac{\tilde \lambda_j}{N}\,=\,-\int_{\tilde {\cal C}} d\tilde y\,\left[K(\tilde x-\tilde y)\,-\,\tfrac{1}{2}K\left(\tilde x-\tilde y- iM\right)\,-\,\tfrac{1}{2}K\left(\tilde x-\tilde y+iM\right)\right]\tilde \rho(\tilde y)-\tfrac{2i\pi\bar \tau}{N}\,\tilde x\,,\\\nonumber\\
&&\hspace{4.6in}\tilde x\in \tilde {\cal C}_j\,,\label{adteq}
\eea
where
\be
K(x)\,=\,\,x\,\ln x^2\,.
\ee
Eq.\eqref{aeq} implies that the mean positions of the individual distributions ${\cal C}_j$ are along the imaginary axis in the $\hat a$-plane.

\subsection{Localization to points with ${\rm Re}(a_D)=0$}
Configurations that extremize the functional ${\cal E}_\tau$ define a genus $N$ Riemann surface in a way that we review in more detail below. This Riemann surface is  the Seiberg-Witten (or Donagi-Witten) curve with the Coulomb branch moduli $\{a_j\}$ specified by A-cycle integrals of the appropriate Seiberg-Witten differential, as illustrated in fig.\eqref{gfig}. The saddle-point equations above ``lock in'' the moduli of the extremizing configurations for the two copies of the instanton partition function that appear in eqs.\eqref{zs4} and \eqref{largevol}.  The Lagrange multipliers $\lambda_j$ also have a natural interpretation in terms of the B-cycle integrals (cf. fig.\eqref{gfig}) of the Seiberg-Witten differential. They are  therefore identified with $a_{Dj}$, the Coulomb branch moduli in the magnetic dual description of the low energy effective theory,
\be
ia_j\,\pm\,\frac{i M}{2}\,=\,\oint_{{\bf A}_j^\pm}dS\,,\qquad\qquad\qquad
a_{D\,j}\,=\, \frac{1}{2\pi i}\left(\lambda_j \,-\,\lambda_{j+1}\right)\,=\,  \oint_{{\bf B}_j}dS\,,
\ee
where $dS$ is the Seiberg-Witten differential:
\bea
&& dS\,=\,\frac{1}{2\pi i}\,x\, 
G(x)\,dx\,,\qquad \qquad \omega(x)\,\equiv\,\int_{C}dy\,\frac{\rho(y)}{x-y}\,,\nonumber\\\nonumber\\
&&G(x)\,\equiv\,\omega\left(x\,-\,\tfrac{iM}{2}\right)\,-\,\omega\left(x\,+\,\tfrac{iM}{2}\right) \,.\label{dS}
\eea
Here $\omega(x)$ is the resolvent function associated to the density $\rho$ and is an analytic function of $x$ with branch-cut singularities along the $N$ intervals $\{{\cal C}_j\}$. By definition, the discontinuity across each branch cut is given by the density function at that point:
\be
\omega(x+i\epsilon)\,-\,\omega(x-i\epsilon)\,=\,-2\pi i\, \rho(x)\,,\qquad
x\in\bigcup_{j}{\cal C}_j\,.
\ee
The function $G(x)$, defined in eq.\eqref{dS}, plays a central role in the solution of the saddle-point conditions and in determining the Riemann surface corresponding to the Donagi-Witten curve. In particular, $G(x)$ has $2N$ branch cuts along the intervals $\{{\cal C}_j\}$ and $\{{\cal C}_j'\}$, as indicated in fig.\eqref{gfig}.
Furthermore, the saddle-point conditions \eqref{adeq} and \eqref{adteq}, when differentiated twice with respect to $x$ (and $\tilde x$), can be recast as  
\be
G\left(x\,-\,\tfrac{iM}{2}\pm i\epsilon\right)\,=\,G\left(x\,+\,\tfrac{iM}{2}\mp i \epsilon\right)\,,  \qquad {x}\in{\cal C}\,.
\ee
These are gluing conditions which identify points immediately above  (below) the cuts $\{{\cal C}_j\}$ with those immediately below (above) the image cuts $\{{\cal C}_j'\}$. This defines a Riemann surface with $N$ handles, whose periods are determined by $\{a_j\}$ and 
$\{a_{D\, j}\}$. This is the Donagi-Witten curve associated to a specific point on the Coulomb branch of ${\cal N}=2^*$ SYM on ${\mathbb R}^4$.

The two remaining saddle-point equations \eqref{lambdaeq} and \eqref{aeq} can now be viewed as $N-1$ independent conditions on the dual periods:
\be
a_{D\,j}\left(ia,\,iM,\,i\tau\right)\,=\,a_{D\,j}\left(-ia,\, -iM,\,-i\bar\tau\right)\,=\,-\,\overline{a_{D\,j}(ia, \,iM,\,i\tau)}\,,\label{adeq1}
\ee
with $ {j=1,2,\ldots N}$. These are precisely the saddle-point conditions we have encountered before, namely,
\be
{\rm Re}(a_{D\,j})\,=\,0\,,\qquad \hat a_j\,=\,ia_j\,.
\ee
These conditions will generically be solved by distributions $\rho(x)$ which may have support in the complex $x$-plane and not necessarily on the real axis alone. As is usual in the steepest descent method, all such saddle-points will have to be summed over and can compete with each other.

\subsection{Point of maximal degeneration}
 
\begin{figure}
\begin{center}
\includegraphics[height=1.9in]{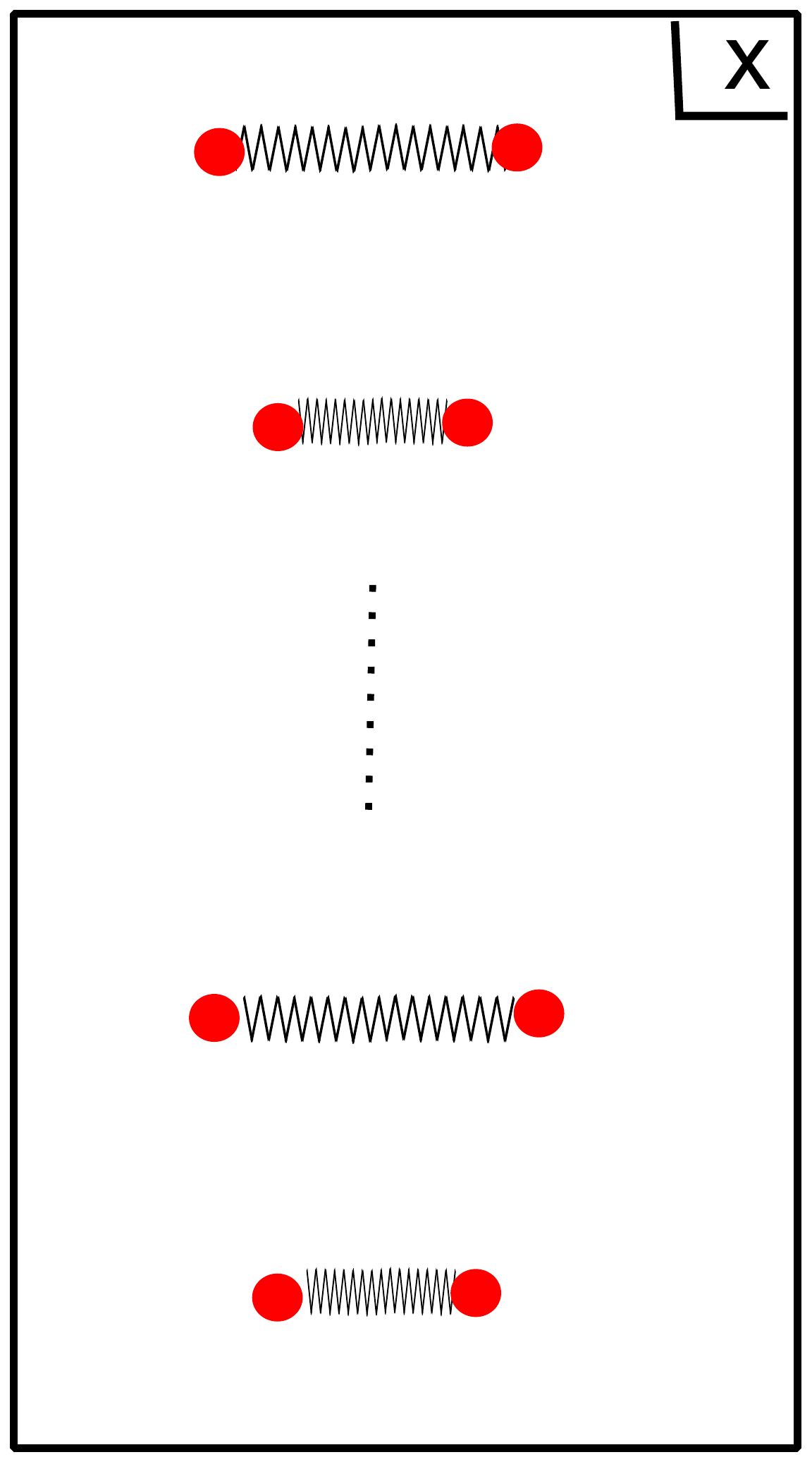}
\hspace{0.9in}
\includegraphics[height=1.9in]{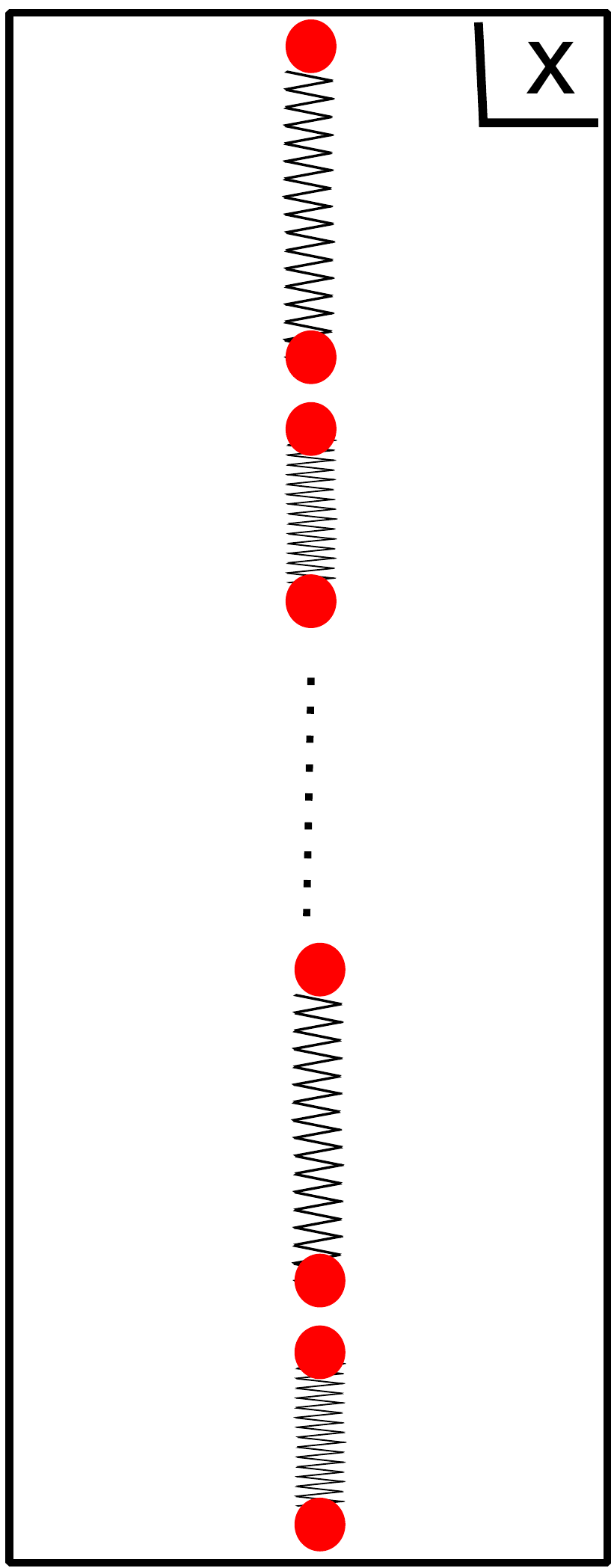}
\hspace{0.9in}
\includegraphics[height=1.93in]{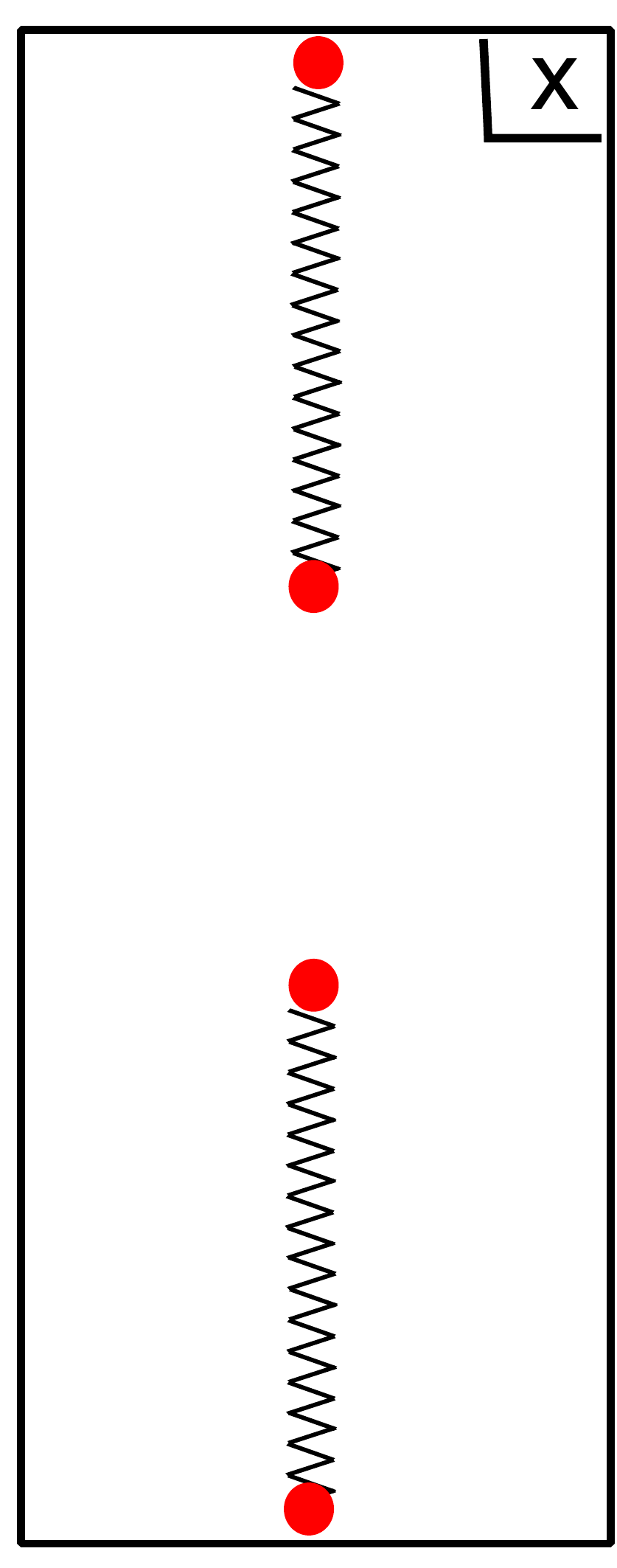}
\end{center}
\caption{\small { {\bf Left:} The $N$ cuts $\{{\cal C}_j\}$ and their images centred at points on the $x$-plane corresponding to purely imaginary periods $\hat a_j=ia_j$. {\bf Centre:} A different orientation of the branch cuts, corresponding to different set of moduli. {\bf Right:} Maximal degeneration of the Donagi-Witten curve to genus one, as the branch cuts of $G(x)$ line up. The two cuts are glued together and all  dependence of physical observables on $N$ enters via the complex structure parameter $\tilde\tau$ of the resulting torus.  
}}
\label{degeneration}
\end{figure}

We look for a saddle-point in a regime where all the cuts ${\cal C}_j$ have extents that are suitably small and the periods satisfy $|a_{jk}|<M$ for all $j,k$. 
Each of the cuts is centred at a point on the imaginary axis in the complex $x$-plane as shown in fig.\eqref{degeneration} (leftmost).  We would now like to understand the maximally degenerate configuration (the dyon singularity) which, as we have argued above, is a saddle-point of the  partition function (for $\theta_{\rm YM}=0$ and $\theta_{\rm YM}=\pi$). Maximal degeneration of the Donagi-Witten curve  occurs when the cuts ${\cal C}_j$ line up end-to-end,  such that end-points of adjacent branch cuts touch each other, as indicated in fig.\eqref{degeneration}. In this limit, $G(x)$ has precisely two branch cuts ${\cal C}$ and ${\cal C}'$ with gluing conditions,\footnote{In the rotated configuration, points to the right(left) of ${\cal C}$ are identified with those to the left(right) of ${\cal C}^\prime$.} yielding a genus one curve. For simplicity, we will {\em assume} that for  imaginary values of the periods $\hat a_j=ia_j$, the cuts ${\cal C}_j$ need to be aligned along the imaginary axis in order for maximal degeneration to occur. This assumption will turn out to be partially justified. We will eventually show that the single branch cut ${\cal C}$, after maximal degeneration of the curve, does lie on the real axis, but only for a finite range of values of the coupling constant.  The branch-points of ${\cal C}$ can move off the imaginary $x$ axis as the coupling constant is increased, while the periods themselves continue to remain purely imaginary.
 
We have already shown that the saddle-point condition ${\rm Re}(a_D)={\rm Re}(a)=0$ can be satisfied at a maximally singular point on the Coulomb branch when $\theta_{\rm YM}=0$ ($N$ even) and $\theta_{\rm YM}=\pi$ ($N$ odd). Therefore, we will proceed with the implicit understanding that the vacuum angle takes one of these two values and interpret our final result in light of this assumption.

In order to calculate the contribution from the maximally degenerate critical point, we first perform the rotations
\be
(x,\, y)\,\to\, (iu,\,iv)\,,
\ee
leaving fixed the normalisation condition
\be
\int_{{\cal C}_j}dx\,\rho(x)\,\to\,\int_{{\cal C}_j}du\,\rho(u)\,=\,\frac{1}{N}\,,
\ee
so that branch cut ${\cal C}$ lies on the real axis in the $u$-plane. This analytic continuation leaves the form of Nekrasov's functional and the ensuing saddle-point equations unchanged. 
Second, since all cuts ${\cal C}_j$ coalesce at such a point, we need only assume that the configuration is characterised by a single branch cut  ${\cal C}$ (and its image under the shift by $M$):
\be
{\cal C}\,=\,\left[-\alpha,\,\alpha \right]\,,\qquad \alpha \in {\mathbb R}\,.
\ee
The requirement that the dual periods have vanishing real parts translates into the equation\footnote{We remark that even if the vacuum angle were non-zero (and equal to $\pi$ for odd $N$), it  would not explicitly appear in the expression for the real part of the dual period.}:
\bea
&&\int_{-\alpha}^\alpha dv\,\rho(v)\left[K(u-v)-\tfrac{1}{2}K(u-v-M)-\tfrac{1}{2}K(u-v+M)\right]\,+\frac{8\pi^2}{g^2_{\rm YM} N}\,u\,=\,0\,,\nonumber\\\label{saddle1}\\\nonumber
&& u\in [-\alpha,\,\alpha]\,.
\eea
Since there is only a single branch cut at a maximally singular point, we do not have immediate access to the values of the individual periods $(a_j, a_{D j})$. To evaluate the $(N-1)$ independent pairs of Seiberg-Witten periods, we would need to move slightly away from the singular point. This is a difficult task for general $N$ and not essential for the immediate problem at hand. 

The remarkable feature of the equation \eqref{saddle1} is that the only dependence on $N$ enters via the term linear in $u$ through the combination,
\be
\lambda\,=\,g^2_{\rm YM}N\,,
\ee
the 't Hooft coupling.
Since we have been consistently working with $N$ fixed, we conclude that the description of the physics at the maximally singular point is {\em large-$N$ exact}. This means that finite $N$ results do not depend separately on $N$ and $g^2_{\rm YM}$, and instead are determined by $\lambda=g^2_{\rm YM}N$.  Therefore, relevant physical observables at such a point are computed exactly by the planar theory. 
This property has been understood in earlier works \cite{mm} within the context of Dijkgraaf-Vafa matrix models \cite{DV} where the planar limit of matrix integrals 
compute  holomorphic sectors of ${\cal N}=1$ SUSY field theories. This applies, in particular, to all the massive vacua of ${\cal N}=1^*$ theory which descend from maximally singular points on the ${\cal N}=2^*$ Coulomb branch.

\subsection{Solution of the saddle-point equation}
We now turn to the solution of the saddle-point equation \eqref{saddle1}.
In order to find the solutions we will closely follow the approach adopted in \cite{mm} for similar matrix integrals which compute holomorphic observables of ${\cal N}=1^*$ theory on ${\mathbb R}^4$. The method is based on the key observation of \cite{kkn} that equations of the type in \eqref{saddle1} can be viewed as specifying a Riemann surface with certain gluing conditions. While this approach was also followed by Russo and Zarembo \cite{zaremborusso2}, we will adopt a slightly different route, placing emphasis on the map from the auxiliary Riemann surface (the degenerate Donagi-Witten curve) to the ``eigenvalue plane'' or the complex $u$-plane. 

\begin{figure}
\begin{center}\includegraphics[width=5.00in]{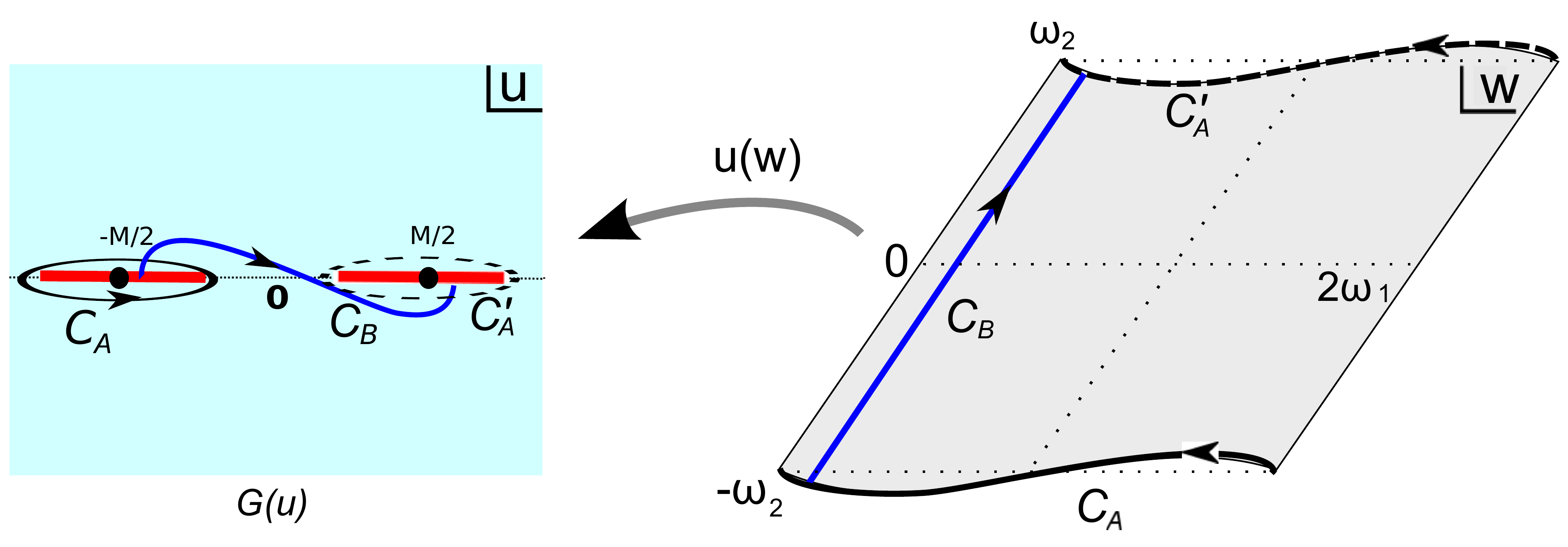}
\end{center}
\caption{\small {The single-cut distribution for the resolvent $\tilde\omega(u)$ leads to an auxiliary torus associated to the function $\tilde G(u)$ satisfying the saddle-point equation. The locations of the branch cuts centred around $u=\pm M/2$ are images of points on the contours $C_A$ and $C_A'$ wrapping the $A$-cycle of the torus. The shaded region in between $C_A$ and $C_A'$ gets mapped to the $u$-plane.}}
\label{torusfig}
\end{figure}

On the $u$-plane ($u=-ix$) we define the resolvent function 
\be
\tilde\omega(u)\,=\,\int_{-\alpha}^\alpha \frac{\rho(v)}{u-v}\,dv\,,\qquad
u\in[-\alpha,\,\alpha]\,,
\ee
It is an analytic function on the complex $u$-plane with a single branch cut singularity on the real axis, the discontinuity across the cut being determined by the density function:
\be
\tilde \omega(u+i\epsilon) \,-\, \tilde\omega(u-i\epsilon)\,=\,-2\pi i \rho(u)\,,\qquad u\in [-\alpha,\,\alpha]\,.
\ee
The resolvent function $\tilde \omega (u)$ on the $u$-plane is related in a simple way to $\omega(x)$ defined on the complex $x$-plane \eqref{dS}, as $\tilde \omega(u)\,=\,i\omega(iu)$
Given the form of eq.\eqref{saddle1}, as before, we introduce the generalised resolvent function:
\be
\tilde G(u)\,=\,\tilde \omega\left(u+\tfrac{M}{2}\right)\,-\,\tilde\omega\left(u-\tfrac{M}{2}\right)\,,\qquad u\in {\mathbb C}\,,\label{deftG}
\ee 
which is now an analytic function of $u$ with {\em two branch cuts} between $[-\alpha +\tfrac{M}{2},\,\alpha +\tfrac{M}{2}]$ and  $[-\alpha -\tfrac{M}{2},\,\alpha-\tfrac{M}{2}]$, with the discontinuities across the cut determined by the density function. For this picture to make sense we must require $\alpha < M/2$, otherwise the two branch cuts of $\tilde G(u)$ would overlap (see fig.\eqref{torusfig}). 
We will explain below that when the extent of the single cut distribution saturates this bound  the branch points of $\tilde G(u)$ move off the real axis into the complex $u$-plane.

Expressed in terms of the generalised resolvent function $\tilde G(u)$, the saddle-point equation becomes
\be
\tilde G\left(u+\tfrac{ M}{2}\pm i\epsilon\right)\,=\,\tilde G\left(u-\tfrac{M}{2}\mp i\epsilon\right)\,,\qquad u\in [-\alpha,\,\alpha]\,,\label{gluesaddle}
\ee
which should be viewed as a gluing condition for the two branch cuts on the $u$-plane. The gluing together of the two branch cuts implies that the auxiliary Riemann surface associated to $\tilde G(u)$ is a torus. Our strategy will be to find the map between the flat coordinates  on this auxiliary torus and the $u$-plane.

\subsubsection{Map from torus to eigenvalue plane}
The auxiliary torus can be viewed as the complex $w$-plane modulo lattice translations, namely ${\mathbb C}_w/\Gamma$ with
\be
\Gamma\,\simeq\,2\omega_1{\mathbb Z}\oplus 2\omega_2{\mathbb Z}\,,\qquad  \qquad \tilde\tau\,=\,\frac{\omega_2}{\omega_1}\,,
\ee 
where we have defined the complex structure parameter $\tilde\tau$ for the torus in terms of its half-periods $\omega_{1,2}$.
The gluing conditions across the two branch cuts imply (see fig.\eqref{torusfig}) that 
\be
u(w+2\omega_1) \,=\, u(w)\,,\qquad u(w + 2\omega_2)\,=\, u(w)\,+ \,M\,.
\ee
Therefore $u(w)$ is a quasi-periodic function on the auxiliary torus with a linear shift under translations by one of the periods. This uniquely fixes $u(w)$ in terms of the Weierstrass $\zeta$-function (see appendix \ref{app:elliptic} for details):
\be
\boxed{u(w)\,=\,i M\,\frac{\omega_1}{\pi }\left(\zeta(w)\,-\frac{\zeta(\omega_1)}{\omega_1}\, w\right)\,.}
\ee
The Weierstrass $\zeta$-function has the property that
\be
\zeta(w + 2\omega_{1,2})\,=\,\zeta(w) \,+\,2\zeta(\omega_{1,2})\,,\qquad \omega_2 \zeta(\omega_1)\,-\,\omega_1\zeta(\omega_2)\,=\,\frac{i\pi}{2}\,.
\ee
It has a simple pole at $w=0$ and its first derivative yields the Wierstrass $\wp$-function:
\be
\zeta'(w)\,=\,-\wp(w)\,,\qquad \zeta(w)\Big |_{w\to 0} \simeq \frac{1}{w}\,+\ldots\,.
\ee
It will also  be useful to re-express $u(w)$ as the logarithmic derivative of the Jacobi theta function,
\be
\boxed{u(w)\,=\,\frac{i M}{2}\,\frac{\vartheta_1'\left(\tfrac{\pi w}{2\omega_1}\right)}{\vartheta_1\left(\tfrac{\pi w}{2\omega_1}\right)}\,.}\label{ujacobi}
\ee
As is customary, without loss of generality we can take one of the periods of the torus to be real:
\bea
2\omega_1\,=\,\pi\,,\qquad 2\omega_2\,=\,\pi\tilde\tau\,.
\eea
The mid-points of the two branch cuts on the $u$-plane at 
$u\,=\,\pm  M/2$ are images of the points $w\,=\,\pm \omega_2$ on the torus:
\be
u(\pm\,\omega_2)\,=\,\pm\frac{M}{2}\,.
\ee
Each of the two branch cuts in the $u$-plane  maps to a separate curve wrapping the $A$-cycle on the auxiliary torus, defined as
\be
{\rm Im}\left[ u(w)\right]\Big|_{w\in C_A}\,=\, 0\,,\qquad\qquad {\rm Im}\left[ u(w)\right]\Big|_{w\in C_A'}\,=\, 0\,. 
\ee
The two curves pass through the points $w=\mp \omega_2$ as sketched qualitatively in fig.\eqref{torusfig}. This condition specifies that the branch cuts on the $u$-plane lie on the real axis. Different choices of orientation of the branch cuts would correspond to different contours on the $w$-plane encircling the $A$-cycle of the torus. 

\subsubsection{The generalised resolvent $\tilde G(u)$}

Our next task will be to find $\tilde G[u(w)]$ as an elliptic function on the $w$-plane i.e. the flat torus. In particular, given the map between the locations of the branch cuts of $\tilde G(u)$ in the $u$-plane and the corresponding curves ${C}_A, C_A^\prime$ in the auxiliary $w$-plane, we have
\be
 \tilde G(u)\, du\big |_{C_A}\,=\,\tilde G\left[u(w)\right]\,u'(w)\,dw\big|_{w\in C_A}\,.
\ee
The function $\tilde G\circ u$, viewed as a function of $w$, must be  doubly periodic i.e. elliptic. This follows from the fact that $\tilde G(u)$ is single-valued when taken around the cycles $C_A$ and $C_B$. From 
the definitions of $\tilde G(u)$ and $u(w)$ we have,
\be
\tilde G(u)\,=\,\tilde G(-u)\,,\qquad \qquad u(w)\,=\,-u(-w)\,,
\ee
which implies that $\tilde G\circ u$ is an {\em even} elliptic function of $w$. Any even elliptic function can be expressed as a rational function of $\wp(w)$ (the Weierstrass $\wp$-function) \cite{ww}. From its definition \eqref{deftG} in terms of the resolvent functions, we deduce the behaviour of $\tilde G$ for large-$u$ (equivalently, $w\to 0$): 
\be
\tilde G(u)\big |_{u \to \infty}\,=\,- M\left(\frac{1}{u^2}\,+\,\frac{M^2\,+\,12 \langle u^2\rangle}{4\,u^4}\,+\ldots\right)\,,
\ee
where we have defined
\be
\langle u^2\rangle\,=\,\int_{-\alpha}^\alpha du\,\rho(u)\, u^2\,.
\ee
Together with the  Laurent expansion of $u(w)$ around $w=0$ (using the identity 
$\zeta(\omega_1)\,=\,E_2/12\omega_1$),
\be
u(w)\big|_{w\to 0}\,=\,
\frac{iM}{2}\left(\frac{1}{w}\,-\,\frac{1}{3}E_2(\tilde \tau)+\ldots\right)\,,
\ee
we obtain the expansion of $\tilde G\circ u$ about $w=0$:
\be
\tilde G[u(w)]\big|_{w\to 0}\,=\,\frac{4 }{
M}\left[w^2\, +\,w^4\left( - 1\,+\, \frac{2}{3}E_2(\tilde\tau)\,-\,\frac{12}{M^2}\langle u^2\rangle\right)+\ldots\right]\,,\label{smallu}
\ee
exhibiting a second order zero at $w=0$. If we assume that $\tilde G[u(w)]$ has no further zeroes in the fundamental parallelogram, then it must have two (simple) poles on the torus \cite{ww}. Therefore, $\tilde G[u(w)]$ can only take the form
\be
\tilde G[u(w)]\,=\,\frac {A}{\wp(w)\,+\,B}\,.\label{G1}
\ee
The coefficients $A$ and $B$ can be fixed by the small $w$ asymptotics of $G[u(w)]$. Comparing the coefficients of the $w^2$ and $w^4$ from \eqref{smallu} and \eqref{G1} in an expansion around $w=0$, we find
\bea
&& A\,=\,\frac{4}{ M}\,,\label{AB}\\\nonumber
&& B\,=\,1\,-\tfrac{2}{3}E_2(\tilde \tau)\,+\,\frac{12}{ M^2}\,\langle u^2\rangle\,.
\eea
The second of these two equations is actually a complicated condition since the right hand side contains 
$\langle u^2\rangle$ which, in principle, itself depends nontrivially  on $B$.  However, we can adopt a shortcut by taking the hint from the observation in \cite{zaremborusso2} that for the saddle-point equation following from eqs. \eqref{saddle1} and \eqref{gluesaddle} the density function $\rho(u)$ necessarily diverges at the end-points of the distribution. The discontinuity of $\tilde G(u)$ is determined by the density $\rho(u)$ and hence $\tilde G(u)$ must diverge at the end-points of the branch cuts.
Since the  Weierstrass $\wp$-function takes every value in the complex plane exactly {\em twice} in the period parallelogram, there are precisely two points in the period parallelogram satisfying the equation $\wp (w)\,=\,-B$ where $\tilde G[u(w)]$ diverges. Labelling the two roots as $w_{1,2}$,
\be
\wp(w_1)\,=\,\wp(w_2)\,=\,-B\,.
\ee
In order for these two points to be identified with end-points of the eigenvalue distribution along the real axis in the $u$-plane, the roots $w_{1,2}$ must lie on ${C}_A$ and $(w_{1,2}+2\omega_2)\in {C_A'}$. Recall that $C_A$ and $C_A'$ are the curves along which $u(w)$ is real. The positions of the two largest eigenvalues (in magnitude) are then determined by the condition,
\be
u'(w)\,=\,-\frac{i M}{2}\left(\wp(w)\,+\,\frac{1}{3}E_2(\tilde \tau)\right)\,=\,0\,,
\qquad
w\in C_A\,,\label{poles}
\ee
which correspond to the extremities of the branch cut on the $u$-plane. Since this equation must have precisely two roots, we must identify them with the poles of $\tilde G[u(w)]$. We conclude that,
\be
\wp(w_{1,2})\,=\,-\frac{E_2(\ttau)}{3}\,= -B\,,\label{BE2}
\ee
and
\be
\boxed{\tG[u(w)]\,=\,\frac{4}{ M}\,\frac{1}{\wp (w)\,+\,\frac{1}{3}E_2(\ttau)}\,=\, -\frac{ 2i}{u'(w)}\,.}
\ee
Crucially, this  formula implies that
\be
\tG(u)\,du\,=\, -2i\, dw\,.
\ee
Its implication is remarkable: {\em Quantum expectation values of physical observables are computed by $A$-cycle integrals on the auxiliary torus with a uniform density function}. In particular, expectation values of single-trace gauge invariant operators, which are given by various moments \cite{Nekrasov:2003rj} of the density function $\rho(u)$ in the $u$-plane, can be expressed in terms of integrals over the $A$-cycle of the torus with uniform density in the $w$-plane:
\bea
\langle u^n\rangle\,&=&\,\frac{1}{2\pi i}\oint_{C_A}dz\,\tG\left(u\right)\, \left(u\,+\,\tfrac{ M}{2}\right)^n
\,=\,-\frac{1}{\pi}\int_{2\omega_1-\omega_2}^{-\omega_2}dw\,\left(u(w)\,+\,\tfrac{ M}{2}\right)^n\\\nonumber\\
&=&\,\frac{i^n M^n}{\pi}\int_{-\frac{\pi}{2}}^{\frac{\pi}{2}}dt\,\left[-\frac{1}{2}\frac{\vartheta_3'(t)}{\vartheta_3(t)}\right]^n\,.\label{moments}
\eea
Since the integrands are analytic functions of $w$, the actual form of the contour is unimportant and the answer only depends on the end-points of the integration range.

 Eq.\eqref{moments} precisely matches previous calculations of condensates at special points on the Coulomb branch of ${\cal N}=2^*$ theory that descend to (oblique) confining vacua of ${\cal N}=1^*$ theory \cite{Dorey:2002ad, mm}. One final step remains in our derivation of the single cut saddle-point of Nekrasov's functional: we have not yet solved for the  modular parameter $\tilde \tau$ of the auxiliary torus. We will address this point below.  Prior to this, we describe a non-trivial consistency check of the solution presented above. Recall that the large-$u$ asymptotics of $\tG$ led us to the condition \eqref{AB} to be satisfied by the constant $B$ which, in turn was determined in \eqref{BE2} by requiring the eigenvalue density to diverge at the end-points of the distribution. These two conditions, when combined, specify the second moment of the eigenvalue distribution:
\be
\langle u^2\rangle\,=\,\frac{M^2}{12}\left(B\,+\,\tfrac{2}{3}E_2\,-\,1\right)\,=\,\frac{M^2}{12}\left(E_2(\tilde\tau)\,-\,1\right)\,.\label{xsquared}
\ee
However, $\langle u^2\rangle$ can also be computed independently using eq.\eqref{moments} and consistency requires that we obtain \eqref{xsquared} via this procedure. Indeed, we find\footnote{We have used the identity
\be
\frac{\vartheta_3'(x)}{\vartheta_3(x)}\,=\,4\sum_{n=1}^\infty\frac{(-1)^n\,\tq^{n/2}}{(1-\tq^n)}\,\sin(2nx)\,,\qquad \tq\,=\,e^{2\pi i \tilde\tau}\,,
\ee
and compared the result of direct integration with the $\tq$-expansion of the Eisenstein series $E_2(\tilde\tau)$.
}
\be
\langle u^2\rangle\,=\,-\frac{M^2}{4\pi}\int_{-\pi/2}^{\pi/2}dt\,\left[\frac{\vartheta'_3(t)}{\vartheta_3(t)}\right]^2\,=\,\frac{M^2}{12}\left(E_2(\tilde\tau)\,-\,1\right)\,.
\ee
This confirms both the  validity of the reasoning used to derive the map $u(w)$ from the torus to the $u$-plane, and the form of $\tG[u(w)]$ that leads to a uniform density function  along the contours  $C_A, C_A^\prime$ on the torus. 

\subsubsection{Fixing $\tilde\tau$ in terms of $\lambda =g^2_{\rm YM}N$}
We can anticipate a constraint on the real part of $\ttau$ by an intuitive argument. Given that the branch cuts in our solution lie on the real axis (at least for some range of $\lambda$) in the $u$-plane, the second moment $\langle u^2\rangle$ must be real and positive. The second Eisenstein series $E_2(\ttau)$ is real when $\tq = e^{2\pi i\tilde\tau}$ is real  (see the $q$-expansion \eqref{qexpansion}). Requiring that $\langle u^2\rangle$ be positive for small $\tq$ (equivalently ${\rm Im}(\ttau) \gg 1$), from eq.\eqref{xsquared} we deduce that 
\be
\tq<0\,\quad\implies\quad {\rm Re}\,\tilde\tau\,=\,\frac{1}{2}\,.
\ee
We will now demonstrate how this constraint and the relationship between $\ttau$ and $\lambda$ emerge naturally from the saddle-point equations. To this end we consider the $B$-cycle integral:
\be
\int_{-\omega_2}^{\omega_2} dw\,=\,2\omega_2\,=\,\pi\ttau\,.
\ee
Using the relation $\tG(u)du\,=\, -2i\,dw$ we rewrite the complex structure parameter $\ttau$ as a $B$-cycle integral on the $u$-plane:
\bea
\int_{\infty}^{u+i\epsilon}\tG\left(v+\tfrac{M}{2}\right)\,dv\,-\,
\int_{\infty}^{u-i\epsilon}\tG\left(v-\tfrac{M}{2}\right)\,dv\,=\,-2i\pi\ttau\,,
\qquad u\in[-\alpha,\,\alpha]\,.\label{intB}
\eea
The integral on the left hand side can be evaluated using the definition of $\tG$ in terms of the resolvent function, keeping track of the imaginary parts following from the $i\epsilon$ prescriptions:
\bea
\int_{\infty}^{u+i\epsilon}\tG\left(v+\tfrac{M}{2}\right)\,dv\,-&&
\int_{\infty}^{u-i\epsilon}\tG\left(v-\tfrac{M}{2}\right)\,dv\,\label{intB2}\\\nonumber
&&=\,-i\pi\,+\,\dashint_{-\alpha}^\alpha
dv\,\rho(v)\,\ln\left[\frac{\left(M^2 -(u-v)^2\right)}{(u-v)^2}\right]\,.
\eea
Now, we note that the integral on the right hand side is constrained by the saddle-point equation \eqref{saddle1}. Differentiating eq.\eqref{saddle1} once with respect to $u$, we obtain
\be
\dashint_{-\mu}^\mu dv\,\rho(v)\,\ln\left[\frac{\left(M^2 -(u-v)^2\right)}{(u-v)^2}\right]\,=\,\frac{8\pi^2}{\lambda}\,,\qquad
u\in[-\alpha,\,\alpha]\,.
\label{saddle3}
\ee
Putting together eqs.\eqref{intB},\eqref{intB2} and \eqref{saddle3}, we finally obtain

\be
\boxed{\tilde\tau\,=\,\frac{4\pi\,i}{\lambda}\,+\,\frac{1}{2}}.\label{ttaurelation}
\ee
Along with the form of the moments \eqref{moments} that compute the condensates at the maximally singular point on the Coulomb branch, this is the second crucial ingredient which forms the basis for the physical interpretation below.

\subsubsection{Physical interpretation of saddle-point}

We now explain in some detail the physical interpretation of the saddle-point obtained above.
The ${\cal N}=2^*$ theory with $SU(N)$ gauge group on ${\mathbb R}^4$ has a family of maximally singular  points at which the genus $N$ Donagi-Witten curve degenerates to a genus one curve. The Donagi-Witten curve is a branched $N$-fold cover of the basic torus with complex structure parameter $\tau$, the complexified microscopic coupling of ${\cal N }=2^*$ theory. At a point of maximal degeneration the curve becomes a torus and is an {\em unbranched} $N$-fold cover of the basic torus with modular parameter $\tau$. An unbranched $N$-fold cover of the basic torus is itself a torus with complex structure parameter $\tilde \tau$ given by  
\cite{dw, Dorey:1999sj, Aharony:2000nt}
\bea
\tilde \tau \,=\, \frac{p\,\tau\,+\, k}{r}\,,\qquad p,r,k \in{\mathbb Z}\,,\label{ttau2}\\\nonumber\\\nonumber
p\,r\,=\, N\,,\qquad k\,=\,0,1,\ldots r-1\,.
\eea
Therefore, the total number of such points is given by $\sum_{r|N} r$, the sum over divisors of $N$. Since the degenerate Donagi-Witten curve at these points is a torus with complex structure parameter $\tilde \tau$, condensates of single trace composite operators,
\be
u_{n}\,=\,\langle{\rm Tr}\Phi^n\rangle\,,
\ee
which are the gauge-invariant coordinates on the Coulomb branch, will naturally be {\em modular} functions of 
${\tilde \tau}$. Modularity follows from $SL(2,{\mathbb Z})$
 transformations on $\tilde \tau$. This duality in the effective coupling $\tilde\tau$, to be contrasted with $SL(2,{\mathbb Z})$ action on $\tau$, was referred to as $\tilde S$-duality in \cite{Aharony:2000nt}.

The saddle-point we have uncovered has complex structure parameter
\be
\ttau\,=\,\frac{{\rm Im}(\tau)}{N}\,+\,\frac{1}{2}\,.
\ee
For even $N$ and $\theta_{\rm YM}=0$, this is the singular point with $p=1$, $r=N$ and $k=N/2$. On the other hand, when $N$ is odd and $\theta_{\rm YM}=\pi$, we can associate this to the singular point with $k=(N-1)/2$. We are now in a position to explain how these precisely match the  physical picture that was anticipated on general grounds in section \ref{sec:N=2*}.

Each maximally singular point on the Coulomb branch corresponds to a distinct supersymmetric vacuum of ${\cal N}=1^*$ theory which is obtained by adding a supersymmetric mass for the adjoint chiral superfield in the ${\cal N}=2^*$ vector multiplet. In a vacuum labelled by an integer  $r$ (which divides $N$ as in eq.\eqref{ttau2}), the $SU(N)$ gauge group is partially Higgsed to $SU(r)$ \cite{vw, Dorey:1999sj}. Classically, the massless fields in such a vacuum constitute an ${\cal N}=1$ vector multiplet with $SU(r)$ gauge symmetry. At low energies these degrees of freedom confine and spawn $r$ discrete vacua (consistent with the Witten index for $SU(r)$, ${\cal N}=1$ SYM) labelled by the integer $k =0, 1,\ldots r-1$. The  massive vacua of ${\cal N}=1^*$ theory are in one to-one correspondence with all possible massive phases of Yang-Mills theory with a ${\mathbb Z}_N$ centre symmetry \cite{dw}. The microscopic $SL(2,{\mathbb Z})$ action on $\tau$ permutes the ${\cal N}=1^*$ phases and therefore, the maximally degenerate points described above. On the other hand, $\tilde S$-duality or the $SL(2,{\mathbb Z})$ action on $\tilde\tau$  is a duality property visible in a given vacuum.

The vacua with $r\,=\,N$ and $k=0,1,\ldots N-1$, are of particular interest to us. These form an $N$-tuplet of confining and oblique confining vacua. The ${\cal N}=1^*$ vacuum labelled by the integer $k$ is associated to the condensation of a dyon with  ${\mathbb Z}_N$-valued magnetic and electric charges $(1,k)$. The oblique confining vacua can be reached from the $k=0$ confining vacuum via shifts of $\theta_{\rm YM}$ by multiples of $2\pi$:
\be
\tau \to \tau\,+\,k\,,\qquad k\,=\,0,1,\ldots N-1\,,
\ee
under which 
\be
\tilde \tau \,=\,\frac{\tau}{N}\,\to\,\frac{\tau\,+\,k}{N}\,.
\ee
In the abelianised description of the ${\cal N}=2^*$ Coulomb branch, the basic confining ${\cal N}=1^*$ vacuum  with $k=0$ descends from the point where $N-1$ BPS-monopoles, carrying magnetic charges under distinct $U(1)$ factors, become massless. This requires the degeneration of $N-1$ independent $B$-cycles of the Donagi-Witten curve. 

The vacuum with $k=N/2$ for $N$ even (and $\theta_{\rm YM}=0$) corresponds to the point with $N-1$  massless BPS dyons carrying charges $(1,N/2)$ under the abelian factors on the Coulomb branch. Analogous statements apply when $N$ is odd and $\theta_{\rm YM}=\pi$. We have therefore confirmed the arguments of section \ref{sec:N=2*}
which picked out these singular points as the saddle-points of the large volume partition function, provided the periods satisfy $a_{ij}<M$.

\subsubsection{Condensates}
The values of the condensates $u_n\,=\,\langle{\rm Tr}\Phi^n\rangle$, which are the gauge invariant coordinates of the point on the Coulomb branch on ${\mathbb R}^4$, are given by the moments \cite{Nekrasov:2003rj} of the eigenvalue distribution \eqref{moments}:
\bea
&& \langle{\rm Tr}\Phi^{2n-1}\rangle\,=\,0\,,\qquad n\in {\mathbb Z}\,,\label{condensates}\\\nonumber\\\nonumber
&&\langle{\rm Tr}\Phi^2\rangle\,=\,N\frac{M^2}{12}\left(1\,-\,E_2(\tilde \tau)\right)\\\nonumber\\\nonumber
&& \langle{\rm Tr}\Phi^4\rangle\,=\,N\frac{M^4}{720}\left[10 E_2(\tilde\tau)^2-E_4(\tilde\tau)-30 E_2(\tilde\tau) +21\right]\,.
\eea
Note that the variables $x$ and $u$ are related as $x=iu$, so that in general  $\langle{\rm Tr}\Phi^n\rangle\,=\,N\langle x^n\rangle=N i^n\langle u^n\rangle$. 
The condensates were already evaluated in earlier works on ${\cal N}=1^*$ theory \cite{mm} and more recently in \cite{zaremborusso1}, and these results are in perfect agreement with eq.\eqref{condensates}.

An important feature of all the condensates is that they are quasi-modular functions of $\tilde\tau$, and therefore possess a
$\tilde q$-expansion or  ``fractional instanton expansion''  since $\tq\,=\,-\exp(2\pi i/g^2_{\rm YM}N)$ \cite{Dorey:2000fc}, which survives the 't Hooft large-$N$ limit.

It is well known that all condensates suffer from scheme dependent, but vacuum independent mixing ambiguities \cite{mm}. The lowest condensate $\langle {\rm Tr}\Phi^2\rangle$ has  an additive ambiguity \cite{dkm}. The dependence on $\ttau$ is, however, vacuum-dependent and physically meaningful, and should be unambiguous. The $\ttau$-dependence and the normalisation of the result above matches the value of $u_{C'}$ for the dyon singularity \eqref{u2su2} in the $SU(2)$ theory which was deduced from the Seiberg-Witten curve. 

\subsubsection{Free energy of the maximally degenerate saddle}
The contribution of the saddle-point to the partition function of the theory on $S^4$ follows directly from the calculation of the second moment $\langle x^2\rangle$ and was also obtained in \cite{zaremborusso1} within the context of the large-$N$ theory. Here we quote the same result which we now know to be valid for any $N$.
 Utilizing the dependence of Nekrasov's partition function on $\tau$, the microscopic gauge coupling, we may write
\be
\frac{\partial \ln {\cal Z}_{S^4}}{\partial\tau_2}\,=\,
2N\pi R^2\, \langle x^2\rangle\,,\qquad \tau_2\,\equiv \,{\rm Im}(\tau)\,.
\ee
This determines the $\tau$-dependent terms in the free energy, and we find,
\bea
&&F\,=\,-\ln{\cal Z}_{S^4}\,=\,-2N^2 R^2 M^2\left(\ln\left|\eta\left(\ttau\right)\right|\,+\,\frac{\pi^2}{3\lambda}\,+\,\frac{1}{2}\ln M\right)\,,\label{lnz}
\\\nonumber\\\nonumber
&&\ttau\,=\,\frac{4\pi i}{\lambda}+\frac{1}{2}\,,\qquad \lambda\,=\, g^2_{\rm YM}N\,.
\eea
The additive coupling-independent piece is fixed by evaluating the action functional on the trivial solution at $g_{\rm YM}=0$. We emphasize that it is not possible to rule out further vacuum-independent (and coupling-dependent) contributions that are a direct consequence of the ambiguity in the condensate $\langle x^2\rangle$. By definition, such ambiguities, which affect the normalisation of the partition function, will not affect the relative free energies between competing saddle points. For the $SU(2)$ theory we have already argued on general grounds that there are no  saddle-points other than the dyon singularity and the free energy of the theory is given by eq.\eqref{lnz} with $N=2$.

It is interesting to examine the behaviour of the free energy of this saddle-point in the strong coupling limit, which could be viewed either as $g_{\rm YM}\gg 1$ for fixed $N$, or as $\lambda \gg 1$ at large-$N$.  The large-$N$ theory has several other saddle points as shown in \cite{zaremborusso1, zarembo} and the maximally degenerate vacuum does not remain a saddle-point for large values of $\lambda$. On the other hand, for the $SU(2)$ theory, we have argued that the dyon singularity is the only saddle-point for all values of $g_{\rm YM}$.  The asymptotic forms of the free energy at small and large couplings are:
\bea
F\,&&=\,-2 N^2 R^2 M^2\,\left(e^{-8\pi^2/\lambda}\,+\,\frac{1}{2}\ln M\,+\ldots\right)\,,\qquad g_{\rm YM} \ll 1\,,\label{lnzexp}\\\nonumber\\\nonumber
&&=\,-2 N^2 R^2 M^2\,\left(-\frac{\lambda}{192}\,+\,\frac{1}{2}\ln \frac{\lambda M}{8\pi}\,+\,e^{-\lambda/8}\,+\ldots\right)\,,\qquad g_{\rm YM}\gg1\,.
\eea
Note that the strong coupling expansion can be taken seriously only for the $SU(2)$ theory where the dyon singularity remains a saddle-point for all values of $g_{\rm YM}$. This is generally not expected to be the case when $N>2$.

It is worth making an important remark at this stage\footnote{We thank the anonymous referee for prompting us to  comment on this.}. For the $SU(2)$ theory the dyon singularity is mapped to itself by the action of S-duality on $\tau$. Under the action of S-duality we have $\tilde\tau\equiv(\tau +1)/2\, \to \tilde\tau'\,\equiv\,(-\tau^{-1} + 1)/2$. The new $\tilde \tau' $ can in turn be mapped back to $\ttau$ by a modular transformation on $\tilde \tau'$ (this is the $SL(2,{\mathbb Z})$-invariance of the degenerate Donagi-Witten curve), namely $\tilde \tau' \to (\tilde \tau' - 1)/(2\tilde\tau'-1)$. If we now assume that the expected S-duality of the  ${\cal N}=2^*$ partition function on $S^4$  should also extend to the limit of large $M$ (or large radius), then the existence of a unique, S-duality invariant saddle point without any phase transitions, points at a consistent picture. The potential manifestation of S-duality of the partition function in the large $M$ limit deserves deeper study. In this context we note that that the $SU(2)$ partition function \eqref{lnz} and its expansions at weak and strong coupling \eqref{lnzexp}, do not precisely exhibit the invariance under S-duality. While the infinite $q$-expansions at weak and strong coupling do map into each other precisely, the term proportional to $1/\lambda$ in \eqref{lnz} and the transformation of the Dedekind-eta function introduce an ``anomaly'', so that the weak and strong coupling expansions are identical only up to these anomalous pieces.

\subsubsection{Increasing $g^2_{\rm YM} N$ and putative non-analyticity}
In our solution for the maximally degenerate saddle-point we began by taking the branch point singularities in Nekrasov's functional to lie on the real axis in te $u$-plane (or imaginary axis in the $x$-plane). This choice was motivated by the purely imaginary values for the periods $\hat a_j\,=\, i a_j$, appearing in Pestun's integral. However, the periods only constrain the integrals of the Seiberg-Witten differential around the cuts or, equivalently, the mean positions  of the individual branch cuts ${\cal C}_j$ (prior to degeneration).  

In the particular case of  the $SU(2)$ theory, we know that the dyon singularity $C^\prime$  lies on the imaginary axis in the $\hat a$-plane and only approaches $\hat a = iM$ asymptotically as $g_{\rm YM}\to \infty$. Therefore this saddle-point cannot exhibit any non-analyticity as a function of $g_{\rm YM}$.
\begin{figure}
\begin{center}
\includegraphics[width=2.65in]{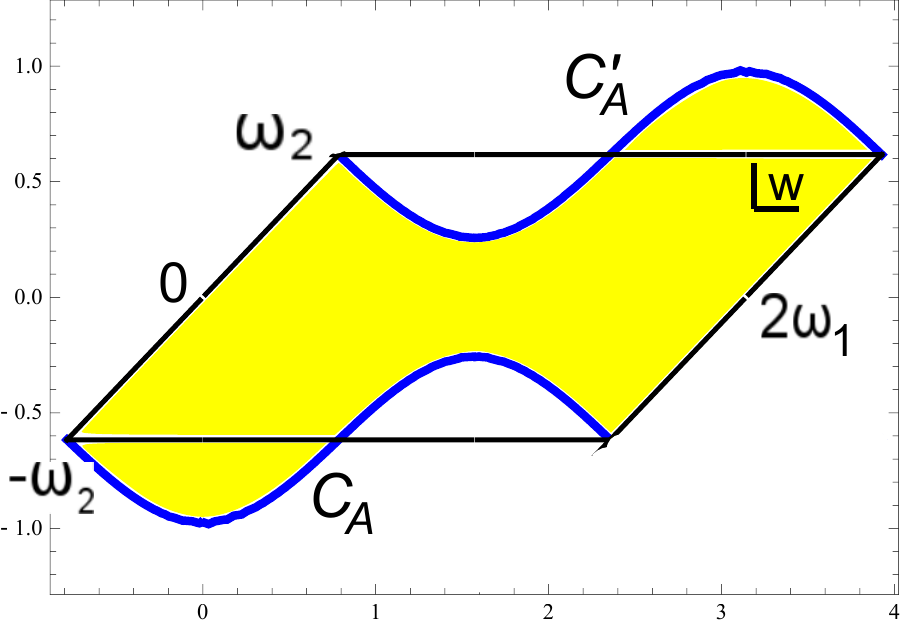}\hspace{0.2in}
\includegraphics[width=2.65in]{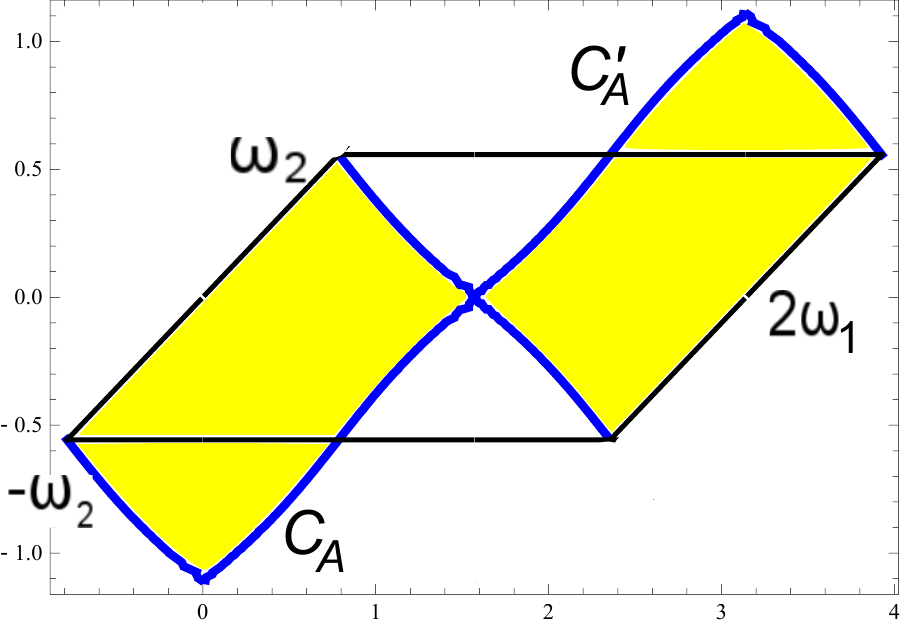}
\end{center}
\caption{\small{{\bf Left:} The shaded region (yellow), isomorphic to the fundamental parallelogram which gets mapped to the $u$-plane for $\lambda\,=\,g^2_{\rm YM}N\,=\,32$. {\bf Right:} The same region, but now at $\lambda \approx 35.45$ when the two branch cuts on the real axis in the $u$-plane touch each other. Above this value of $\lambda$ the branch points move off the real axis into the complex plane.}}
\label{critminus}
\end{figure}
 On the other hand, the positions of the branch cuts of $\tG(u)$ on the $u$-plane in fig.\eqref{torusfig} suggest that it is possible for the end-points of the branch cuts to touch each other at $u=0$ for some value of $g_{\rm YM}$, posing a possible source of non-analyticity when this happens. We now explain the implication of this phenomenon.
 
\begin{figure}
\begin{center}
\includegraphics[width=2.65in]{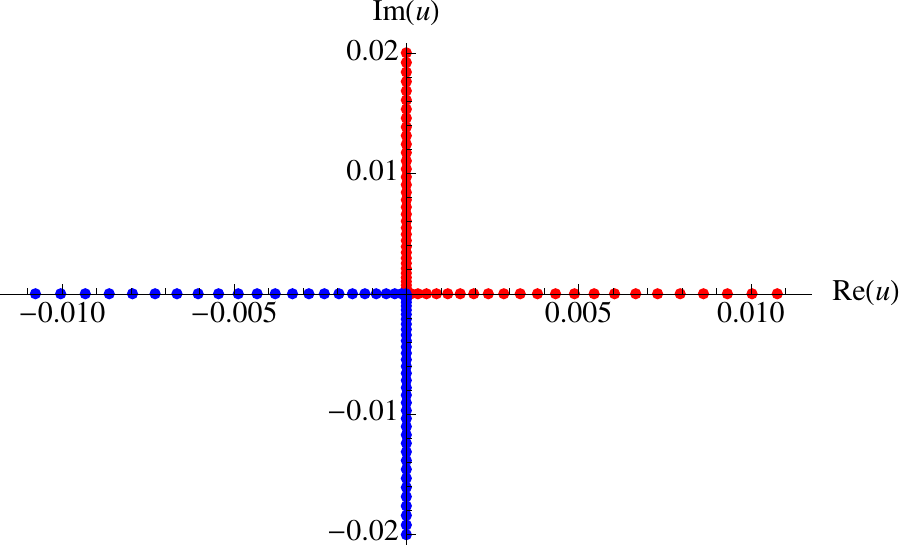}
\end{center}
\caption{\small{The positions of the two branch point singularities in fig.\eqref{torusfig}, closest to $u=0$, are plotted in blue and red, for different values of $\lambda$ in the range $32 \leq \lambda\leq 39$. Near $\lambda\approx 35.45$, the two branch points collide at $u=0$ and then move off along the imaginary axis in the $u$-plane.}}
\label{branchpt}
\end{figure}

The density function $\rho(u)$ diverges at the end-points of the branch cuts. This is reflected in the fact that the resolvent function $\tG[u(w)]$ has two simple poles 
on the  curve $C_A$ in the $w$-plane, the locations of which (and their translates by $2\omega_2$ on $C_A^\prime$) correspond to the branch points in the $u$-plane. Therefore, if the two branch cuts in fig.\eqref{torusfig} were to meet at $u=0$, this would be signalled by the appearance of a double pole for $\tG[u(w)]$ on the torus. Since $\tG[u(w)]\sim (\wp (w) +\tfrac{1}{3}E_2)^{-1}$, this implies a double-zero for $(\wp(w)+\tfrac{1}{3}E_2)$ at $u(w)=0$.  A double pole in $\tG[u(w)]$ appears when  $\wp'(w)$ vanishes, i.e. at the half-periods of the torus where $\wp'(\omega_1)\,=\,\wp'(\omega_2)\,=\,\wp'(\omega_1+\omega_2) =0$. Noting that the origin $u=0$ corresponds precisely to the  half-period, $w=\omega_1=\pi/2$, we expect the two branch cuts in the $u$-plane to collide at the origin when
\bea
&&\wp\left(\left.\frac{\pi}{2}\,\right|\,\tilde\tau\right)\,=\,-\frac{1}{3}\,E_2(\tilde\tau)\,.
\eea
Making use of the identity $\wp\left(\tfrac{\pi}{2}\right)\,=\,\tfrac{2}{3}\left(\,2E_2(2\tilde\tau)\,-\,E_2(\tilde\tau)\right)$, this can be recast as a condition on the complex structure parameter,
\be
E_2(\tilde\tau)\,=\,4 E_2(2\tilde\tau)\,,
\ee
which is satisfied when $\lambda\approx 35.5$. Fig.\eqref{critminus} depicts what happens to the cycles $C_A$ and $C_A^\prime$ on the auxiliary torus at this value of the gauge coupling. Beyond this value of the gauge coupling the branch-points simply move off the real axis (see fig.\eqref{branchpt})\footnote{The locations of the branch points on the $u$-plane can be determined by first solving for points $w_*$ on the torus that satisfy $u'(w_*)=0$, corresponding to the extrema of $u(w)$ and then 
obtaining the coordinates of these points on the $u$-plane i.e. $u(w_*)$.}.
At finite $N$, and for the specific case of $N=2$, this phenomenon does not translate into a non-analyticity of the partition function.
Clearly, the free energy of the degenerate saddle-point does not exhibit  non-analyticities at this value of $\lambda$. This is  confirmed by our observations on the $SU(2)$ theory where motion of the branch points into the complex plane cannot affect the period $\hat a$, which remains purely imaginary at arbitrarily large values of $g_{\rm YM}$ (approaching $\hat a = iM$ asymptotically).

In the large-$N$ limit however, the situation is different. We will elaborate on this below.
 Although the characterisation of the saddle-point is identical to the finite $N$ case, the motion of the branch points into the complex $u$-plane simultaneously implies that some of the periods $\hat a_j$ at the maximally singular point move off into the complex $u$-plane. This takes the singular point away from Pestun's contour ${\rm Re}(\hat a_j)\,=\,0$. Therefore, beyond this critical value of the 't Hooft coupling in the large-$N$ theory, the large radius partition function should be computed by a different saddle-point, as was concluded in 
 \cite{zaremborusso1,zaremborusso2,zarembo}.
 
 Based on our observations on the $N=2$ and $N=\infty$ cases, we may conclude that for any fixed $N$, the meeting of the branch points of $\tG$, and their subsequent motion off the axis of real $u$, does not by itself imply non-analyticity in the partition function. A putative non-analyticity can be expected when some of the periods at the saddle-point configuration cease to be purely imaginary. From our analysis in section \ref{sec:saddles}, this is likely to occur when one of the differences $\hat a_{j,j+1}$ approaches $iM$ and a massless hypermultiplet appears. For theories with $N>2$, this is a collision of the dyon singularity with a singularity which is not maximal, but has a massless electric hypermutiplet \footnote{ Points of maximal degeneration on the Coulomb branch (or massive vacua of ${\cal N}=1^*$ theory) are permuted in a definite fashion by the microscope $SL(2,{\mathbb Z})$ duality of ${\cal N}=4$ SYM. They cannot merge or collide \cite{dw}, as already seen for the $SU(2)$ theory. In particular, this means that the dyon singularity (for any $N$) cannot collide with the singularity where $(N-1)$ massless electric hypermultiplets appear.}.  This leads to an Argyres-Douglas type singularity. Beyond a critical value of the gauge coupling $\lambda_c(N)\equiv \left(g^2_{\rm YM}\right)_c N$ we should expect the partition function to be computed by a new saddle point of the `mixed' type as discussed in section \ref{sec:saddles}. For any fixed $N$, we further expect $\lambda_c (N)$ to be larger than the value of $\lambda$ at which the branch-points of $\tG$ collide ($\lambda\approx 35.45$).

\section{Large $N$ vs finite  $N$} 
We have shown that one solution to the large volume saddle-point equations following from Nekrasov's partition function is a point of maximal degeneration on the Coulomb branch satisfying ${\rm Re}(a_{D\,j})\,=\,{\rm Re}(\hat a_j)\,=\,0$. It has been known for sometime that gauge-invariant observables at such points are large-$N$ exact \cite{mm}, meaning that, up to vacuum-independent mixing ambiguities they are computed by planar graphs at finite $N$. This is manifest from the viewpoint of  the Nekrasov action functional (at large $R$)  since the relevant saddle-point is obtained by the merger of $N$ branch cuts $\{{\cal C}_j\}$ into a single cut, and all dependence on $N$  is in the effective modular parameter ${\rm Im}\tilde\tau = 4\pi i /g^2_{\rm YM}N$.  Therefore, the density function $\rho(x)$ on this branch cut, the condensates and free energy depend only on the 't Hooft coupling, and their functional forms  are unaffected by $N$ (at this saddle-point).

The $N-1$ periods and their duals can be accessed by moving infinitesimally away from the single-cut configuration. The $N$-dependence of the theory is encoded non-trivially in the Seiberg-Witten periods at generic points in the moduli space. When the gauge coupling is small ($g^2_{\rm YM} \ll 1$), the extents of the cuts ${\cal C}_j$ can be made parametrically small, and the branch cuts can be be replaced by point-like singularities, 
\be
\rho(x)\,\to\ \frac{1}{N}\sum_{j=1}^{N}\delta(x\,-\, i a_j)\,,
\ee 
where one must also require the separations between the cuts $\sim | a_j-  a_k|$ to be relatively large. In this limit  Nekrasov's functional \eqref{action} reduces to the one-loop prepotential:
\bea
{\cal E}_{\tau}[\rho]&&\to\label{limiting}\\\nonumber
&&\frac{1}{4}\sum_{j,k}\left[a_{jk}^2\ln a_{jk}^2\,-\,\frac{1}{2}(a_{jk} - M)^2\ln(a_{jk} - M)^2\,-\,\frac{1}{2}(a_{jk} + M)^2\ln(a_{jk} + M)^2\right]\\\nonumber
\\\nonumber
&& +\,\frac{4\pi^2}{g^2_{\rm YM}}\sum_{j} a_j^2\,.
\eea
For finite values of $N$, this approximation will break down when the mean positions of the cuts are close to each other, i.e. the differences $|a_{jk}|$ are comparable to the extents of the cuts ${\cal C}_j$. This is when instanton contributions can no longer be ignored. On the other hand, in the 't Hooft large-$N$ limit, when instantons are exponentially suppressed ($\sim e^{-N})$, the above picture should become exact. The saddle points of the large-$N$ functional, which is now precisely the one-loop prepotential, can be used to examine saddle points in a large-$N$ continuum picture:
\be
\frac{1}{N}\sum_{j\neq k} \,\to\, \dashint_{-\alpha}^\alpha du\, \hat\rho(u)\,,\qquad \int_{-\alpha}^\alpha du\,\hat\rho(u)\,=\,1\,.
\ee
$\hat \rho(u)$ can be interpreted as  a large-$N$ density for the periods $a_j$, which can also be viewed as the eigenvalues of the adjoint scalar $\Phi$. Crucially, the large-$N$ saddle point equation for the continuum distribution of the 
eigenvalues is,
\bea
&&\dashint_{-\alpha}^\alpha dv\,\hat\rho(v)\,\left[\frac{1}{2}K(u-v+M)\, +\,\frac{1}{2} K(u-v-M)\,-\, K(u-v)\right]\,=\, \frac{8\pi^2}{\lambda}\,u\,.\nonumber\\\nonumber\\
&& u\in[-\alpha,\alpha]\,,\qquad K(x)\,=\, x\,\ln(x^2)\,.\label{saddleN}
\eea
This is the saddle-point equation \eqref{saddle1} that we have already solved at fixed $N$, and all results of section \ref{sec:nekrasov} automatically apply without any changes. The one crucial difference is in the interpretation of the density function $\hat \rho(u)$ which now represents the large-$N$ distribution of the eigenvalues $a_j$. 

The works of \cite{zaremborusso1, zaremborusso2} analysed  this matrix model and its saddle-points, by directly taking the large-$R$ limit of Pestun's matrix integral as originally presented in \cite{pestun}:
\bea
{\cal Z}_{S^4}\,=\,\int d^{N-1}a\,\prod_{i < j}\,
\frac{(a_i-a_j)^2\,H^2(a_i-a_j)}{H(a_i-a_j +M)\, H(a_i - a_j -M)}\,\,e^{-\frac{8\pi^2\,R^2}{g^2_{\rm YM}}\sum_k a_k^2}\,|{\cal Z}_{\rm inst}|^2\,
\label{mmZ}
\eea
where the function $H(x)$ is defined as
\be
H(x)\,\equiv\,\prod_{n=1}^\infty\left(1\,+\,\frac{x^2R^2}{n^2}\right)^n\,e^{-x^2 R^2/n}\,.
\ee
It can intuitively be understood as a combination of one-loop fluctuation determinants obtained by integrating out heavy modes around a Coulomb-branch like configuration of VEVs, and non-perturbative instanton contributions. In the limit $MR \gg 1$, the integrand can be expressed as the exponential of \eqref{limiting}, assuming that instantons can be neglected in the large-$N$ limit.

\subsection{The distribution of periods at large-$N$}
When the eigenvalues of the large-$R$ matrix integral condense on a single branch cut, we obtain the maximally degenerate saddle-point. Given that the saddle-point equations are identical, our analysis in section \ref{sec:nekrasov} directly yields the locations of the periods $\{\hat a_j\,=\,ia_j\}$ on this branch cut. In particular, their positions, when mapped to the auxiliary torus (the degenerate Donagi-Witten curve), are uniformly distributed along the cycles $C_A$ and $C_A'$ in the $w$-plane, shown in fig.\eqref{torusfig}. The values of the periods i.e. their locations in the $x$-plane $(x= iu)$ are given in the large-$N$ continuum limit as (using eqs.\eqref{ujacobi} and \eqref{logtheta}) 
\be
\hat a(w)\,=\,iu(w+\omega_2)\,=\,-\frac{M}{2}\,\frac{\vartheta_3'\left(w-\tfrac{\pi}{2}\right)}{\vartheta_3\left(w-\tfrac{\pi}{2}\right)}\,,\qquad (w+\omega_2)\in C_A'\,.
\label{laxformula}
\ee
The parameter $w$ should be viewed as a continuous label parametrising the points on the cycle $C_A'$ on the torus
. This result for the large-$N$ eigenvalues $\hat a$ at the $\left(1,\left[\tfrac{N}{2}\right]\right)$ dyon singularity coincides with the exact formula for the eigenvalues of the adjoint scalar (in the ${\cal N}=2$ vector multiplet) in the massive vacua of ${\cal N}=1^*$ theory, derived sometime ago in \cite{Dorey:2002ad}. In that paper, the connection between supersymmetric gauge theories and integrable systems \cite{dw, Martinec:1995by, Gorsky:1995zq} was exploited to relate the adjoint scalar eigenvalues directly to the eigenvalues of the Lax matrix at equilibrium positions of the $N$-body elliptic Calogero-Moser Hamiltonian. We note that it is only in the large-$N$ limit that we are able to identify the periods with eigenvalues of the adjoint scalar (see also \cite{Douglas:1995nw}). At any finite $N$ they are distinct observables, not to be confused with each other.

At weak 't Hooft coupling $\lambda=g^2_{\rm YM}N \ll 1$,  eq.\eqref{laxformula} can be simplified using the identity \eqref{thetaq},
\be
\hat a( t)\to -2i M\,e^{-4\pi^2/\lambda}\,\sin(2t)\,,\qquad t\in\left[
-\tfrac{\pi}{2},\,\tfrac{\pi}{2}\right]\,.
\ee
Therefore, in this limit, $\hat a$ is purely imaginary as required and the density function $\hat \rho(\hat a=i a)\,\sim\,1/\sqrt{\Lambda^2 - a^2}$, matching the result of Douglas and Shenker \cite{Douglas:1995nw} for the pure ${\cal N}=2$ theory at large-$N$. 

At strong 't Hooft coupling $\lambda\gg 1$, the eigenvalues at the singular point can be evaluated after performing a modular transformation \eqref{thetamodular} which yields
\be
\hat a (t)\to \frac{M\lambda}{8\pi^2}\,t\,,\qquad t\in \left[
-\tfrac{\pi}{2},\,\tfrac{\pi}{2}\right]\,.
\ee
At arbitrarily strong coupling, the values of the periods $\hat a$ at this point are all {\em real} with a uniform density. Therefore, the maximally singular point does not lie on Pestun's contour of integration ${\rm Re}(\hat a)\,=\,0$. This is in line with our previous observation and that of \cite{zaremborusso1, zaremborusso2} that at large-$N$, when $\lambda$ is dialled beyond the critical value $\lambda_c\approx 35.45$, the theory should undergo a phase transition to a new saddle point\footnote{In fact, it is possible to exclude all points on the Coulomb branch that are maximally singular with $\tilde \tau \,=\, (p\tau+k)/r$ where $r\sim {\cal O}(N)$ and $p\,r\,=N$.
In the limit of strong 't Hooft coupling they all have  periods that are real-valued  (and uniformly distributed) in this limit.}. Therefore the new saddle-point cannot be a point of maximal degeneration. Indeed, as explained in section \ref{sec:saddles}, when some of the period differences exceed $M$, the saddle-point conditions are non-trivial and are not conditions for the maximal degeneration of the Donagi-Witten curve.

It has been argued in \cite{zaremborusso1, zaremborusso2} that, following an infinite sequence of phase transitions, when $\lambda\gg MR \gg 1$, the large-$N$ saddle point which dominates the partition function is remarkably simple and given by the solution to Gaussian matrix model, namely the Wigner eigenvalue distribution. The intuitive argument for this follows by assuming that the extent of the distribution at strong coupling has is controlled by $\lambda$ which is taken to be much larger than $M$. Then, the one-loop prepotential can be formally expanded for small $M$ to yield 
\be
{\cal F}_{\rm 1-loop}\to -\frac{M^2}{2}\sum_{j,k}\ln(a_j-a_k)^2\,+\,\frac{8\pi^2}{\lambda}N\sum_{j} a_j^2\,.
\ee
The large-$N$ saddle-point is characterised by the Wigner semicircle distribution
\be
\hat \rho(a)\,=\,\frac{8\pi}{\lambda M^2}\sqrt{\frac{\lambda M^2}{4\pi^2}\,-\,a^2}\,,
\ee
and free energy 
\be
F\to -\frac{1}{2}N^2M^2R^2\ln\frac{\lambda M^2}{8\pi^2}\,.
\ee
Although it is possible to find maximally degenerate vacua of the large-$N$ theory which exhibit the same scaling of the free energy with $\lambda$, the scaling of the eigenvalue distributions and the moments with $\lambda$ cannot be reproduced by such vacua. The Wigner distribution implies that the condensates must scale as $\langle{\rm Tr}\Phi^{2n}\rangle\sim N M^{2n}\lambda^n$ in the large $\lambda$ limit. On the other hand, for the $N$-tuplet of maximally degenerate singularities (corresponding to the confining and oblique confining phases of ${\cal N}=1^*$ SYM), a straightforward strong coupling limit yields the scaling $\langle{\rm Tr}\Phi^{2n}\rangle\sim N M^{2n}\lambda^{2n}$.

\section{Discussion}
In this paper we have obtained a complete characterisation of a particular (maximally degenerate) saddle-point of the partition function of ${\cal N}=2^*$ theory on a large $S^4$ for fixed $N$. For the $SU(2)$ theory this is sufficient to compute the partition function, while for higher rank gauge groups we also need to quantify the contributions from other saddle-points. We have outlined the criteria and conditions to be satisfied by the Seiberg-Witten periods at such additional saddle-points. The immediate outstanding question is whether the `cuspy' configurations studied in detail at large-$N$ in \cite{zaremborusso1, zaremborusso2} can be shown to correspond to the general category of saddle-points we have discussed.

We have argued in general that when the maximally singular saddle point approaches (with increasing coupling strength) a point on the integration contour where an electric hypermultiplet becomes light, the point of maximal degeneration ceases to be a saddle-point and, subsequently, moves off into the complex plane. We can trace the origin of this phenomenon to the fact that the prepotential is a function with branch point singularities and when a saddle-point approaches such a point, we expect non-analytic behaviour of some sort. It would be extremely interesting to understand this phenomenon in detail for the $SU(3)$ theory.

We have focussed attention exclusively on the critical points of the function ${\cal F}(ia)+{\cal F}(-ia)$ for real $a$. However, as is customary in the saddle-point method, one must also look at critical points in the complex plane which can contribute to the integral in question following smooth deformation of the integration path so that it passes through such complex saddle points.  It is important to understand whether such complex saddle points exist for the Pestun partition function (at large volume)  and whether they can compete with the saddle-points already discussed in this paper and in previous works  \cite{zaremborusso1, zaremborusso2}. One of the puzzles that this may also potentially shed light on is the question of S-duality of the ${\cal N}=2^*$ partition function on $S^4$ for $N>2$. S-duality on $\tau$ maps the dyon singularity (which is the low-$\lambda$ saddle point) to a maximal singularity where $\ttau = N\tau/4 +1/2$.  This latter singular point would descend to an ${\cal N}=1^*$ vacuum where $SU(N)$ is first Higgsed to an $SU(2)$ which then confines.
Such points correspond to  specific multi-cut solutions of the Nekrasov ``matrix model''. However such a point also does not appear to satisfy the saddle-point equations we have discussed for imaginary $\hat a_i$ (or real $a_i$). How $S$-duality manifests itself in the large radius limit and the potential role played by complex saddle-points is a very interesting and important issue  for a complete understanding of the partition function.

\acknowledgments The authors would like to acknowledge partial financial support from the UK  Science and Technology Facilities Council (STFC) grant no. ST/L000369/1.

\newpage
\appendix
\section*{Appendix}
\section{Elliptic functions and modular forms}
\label{app:elliptic}
We provide some details of the properties of elliptic, quasi-elliptic functions and modular forms that are useful for our calculations. For a more complete treatment we refer the reader to \cite{ww} and \cite{koblitz}. 

\subsection{Elliptic and quasi-elliptic functions}
An elliptic function is a function on the complex plane, periodic with two periods $2\omega_1$ and $2\omega_2$. Defining the lattice $\Gamma\,=\,2\omega_1{\mathbb Z}\oplus2\omega_2{\mathbb Z}$ and the basic period parallelogram as
\be
{\cal D}\,=\,\left\{z=2\mu\omega_1+2\nu\omega_2\,|\,\mu,\nu\in[0,1)\right\}\,,
\ee
the Weierstrass $\wp$-function is analytic in ${\cal D}$ except at $z=0$, where it has a double pole. ${\wp}(z)$ is an even function of $z$, defined via the sum
\be
\wp(z;\omega_1,\omega_2)\,=\,\frac{1}{z^2}+\sum_{(m,n)\neq(0,0)}\left\{\frac{1}{(z-2m\omega_1-2n\omega_2)^2}-\frac{1}{(2m\omega_1+2n\omega_2)^2}\right\}\,.
\ee
\begin{itemize}
\item{The Weierstrass $\wp$-function satisfies
\be
\wp^{\prime}(z)^2\,=\,4\wp(z)^3-g_2\wp(z)-g_3\,;\qquad
g_2\,=\,60\sum\Omega_{mn}^{-4}\,,\quad
g_3\,=\,140\sum\Omega_{mn}^{-6}\,,
\ee
where $\Omega_{mn}=2m\omega_1+2n\omega_2$ with $(m,n)\neq (0,0)$ and $g_2, g_3$ are the {\em Weierstrass invariants}.} 

\item{Under a modular transformation, of the complex structure parameter $\tau\,=\,\omega_2/\omega_1$ of the torus, the $\wp$-functions transforms with weight two:
\be
\wp\left(\frac{z}{\tau};\, \omega_1,\,-\frac{\omega_1}{\tau}\right)\,=\,
\tau^2\,\wp(z;\, \omega_1,\,\tau\omega_1)\,.\label{pmodular}
\ee}
\item{The following formula can be used to infer the behaviour of $\wp(z)$ in the limit
${\rm Im}(\tau)\gg 1$ \cite{ww}
\be
\wp(z;\omega_1,\omega_2)\,=\,\left(\tfrac{\pi}{2\omega_1}\right)^2\left[-\tfrac{1}{3}\,+\,\sum_{n=-\infty}^\infty{\rm csc}^2\left(\tfrac{z-2n\omega_2}{2\omega_1}\pi\right)\,-\,\sum_{n=-\infty}^{\infty\prime}{\rm csc}^2\left(\tfrac{n\omega_2}{\omega_1}\pi\right)\right].\label{plowt}
\ee}
\item{The Weierstrass $\zeta$-function is a quasi-elliptic function, analytic in ${\cal D}$ but with a simple pole at $z=0$ 
\bea
&&\wp(z)\,=\,-\zeta'(z)\,,\qquad\qquad\zeta(z+2\omega_{1,2})\,=\,\zeta(z)+2\zeta(\omega_{1,2})\label{zetadef}\label{zetasigmadef}
\eea
}

\item{A slight modification renders the Weierstrass $\zeta$-function periodic along one of the periods of the torus ${\mathbb C}/\Gamma$
\be
\tilde\zeta(z)\,=\,\zeta(z)\,-\,\frac{\zeta(\omega_1)}{\omega_1}\,z\,.
\label{zetatdef}
\ee
The new function is periodic along the A-cycle, but only quasiperiodic along the B-cycle.
\be
\tilde\zeta(z+2\omega_1)\,=\,\zeta(z)\,,\qquad\zeta(z+2\omega_2)\,=\,
\zeta(z)-\frac{i\pi}{\omega_1},
\ee
where we have used the identity
\be
\omega_2\zeta(\omega_1)-\omega_1\zeta(\omega_2)\,=\,i\pi\,.
\ee}
\item{The quasiperiodic function $\tilde\zeta(z)$ is the logarithmic derivative of the Jacobi theta function
\be
\tilde\zeta(z)\,=\,\frac{\pi}{2\omega_1}\,\frac{\vartheta_1'\left(\frac{\pi z}{2\omega_1}\right)}{\vartheta_1\left(\frac{\pi z}{2\omega_1}\right)}\,
\equiv\, -\frac{i\pi}{2\omega_1}\,+\,\left.\frac{\pi}{2\omega_1}\,\frac{\vartheta_3'\left(\frac{\pi z'}{2\omega_1}\right)}{\vartheta_3\left(\frac{\pi z'}{2\omega_1}\right)}\right|_{z'=z-\omega_1-\omega_2}.
\label{logtheta}
\ee
}
\item{The Jacobi-theta functions satisfy the heat equation:
\be
\frac{\partial^2\vartheta_\nu(u|\tau)}{\partial u^2}\,=\,-\frac{4}{i\pi}\frac{\partial\vartheta_\nu(u|\tau)}{\partial\tau}\,.\label{heateq}
\ee
}
\item{The logarithmic derivative of $\vartheta_3$ has a useful $q$-expansion 
\be
\frac{\vartheta_3'(z)}{\vartheta_3(z)}\,=\,-4\sum_{n=1}^\infty\frac{q^{n-1/2}\sin 2z}{1\,+\,2q^{n-1/2}\cos 2z\,+\,q^{2n-1}}\label{thetaq}
\ee
where $q=e^{2\pi i \tau}$.
 }
\item{Under the special modular transformations $\tau \to \tau +2$ and $\tau\to -1/\tau$, 
\be
\vartheta_3(z,\tau+2)\,=\,\vartheta_3(z,\tau)\,,\qquad
\vartheta_3\left(\tfrac{z}{\tau},-\tfrac{1}{\tau}\right)\,=\,\sqrt{-i\tau}\,e^{-z^2/i\pi\tau}\,\vartheta_3(z,\tau)\,,\label{thetamodular}
\ee

}
\end{itemize}
\subsection{The Eisenstein series }
There are a number of ways to introduce the Eisenstein series (see \cite{koblitz})
\bea
E_k(\tau)\,=\,\frac{1}{2}\sum_{\substack{m,n \in{\mathbb Z}\\(m,n)=1}}\frac{1}{(m\tau+n)^k}\,,\qquad\qquad{\tau}=\frac{\omega_2}{\omega_1}
\eea
where $\tau$ is the complex structure parameter of the torus defined by ${\mathbb C}/\Gamma$ and $(m,n)$ denotes the greatest common divisor. Each series has a $q$-expansion
\bea
&&E_2(\tau)\,=\,1-24\sum_{n=1}^\infty\sigma_1(n)\,q^n\,,\qquad q=e^{2\pi i \tau}
\label{qexpansion}\\\nonumber
&&E_4(\tau)\,=\,1+240\sum_{n=1}^\infty\sigma_3(n)\,q^n\,,\\\nonumber
&&E_6(\tau)\,=\,1-504\sum_{n=1}^\infty\sigma_5(n)\,q^n\,,
\eea
where $\sigma_j(n)$ is a sum over each positive integral divisor of $n$ raised to the $j^{\rm th}$ power. Under the S-transformation $\tau\to -1/\tau$, the modular forms with the exception of $E_2(\tau)$, transform covariantly with a specific modular weight
\be
E_k(-1/\tau)=\tau^k\,E_k(\tau)\qquad k\geq 4\,,\qquad E_2(-1/\tau)=\tau^2E_2(\tau) +\frac{6\tau}{i\pi}\,.\label{modularE2}
\ee
The anomalous transformation of $E_2(\tau)$ can be fixed by a shift.
\be
\widehat E_2(\tau,\bar\tau) = E_2(\tau)-\frac{3}{\pi {\rm Im}(\tau)}\,.
\ee
This is a modular form of weight two, although it is not holomorphic. Further useful relations include
\be
\zeta(\omega_1)=\frac{\pi^2}{12\omega_1} E_2(\tau)\,,\qquad
g_2=\frac{4\pi^4}{3(2\omega_1)^4}E_4(\tau)\,,\qquad
g_3\,=\,\frac{8\pi^6}{27 (2\omega_1)^6}\,E_6(\tau)\,.
\ee


\newpage
\begin{thebibliography}{99} 

\bibitem{pestun} 
  V.~Pestun,
  ``Localization of gauge theory on a four-sphere and supersymmetric Wilson loops,''
  Commun.\ Math.\ Phys.\  {\bf 313}, 71 (2012)
  [arXiv:0712.2824 [hep-th]].

\bibitem{drukkergross} 
  N.~Drukker and D.~J.~Gross,
  ``An Exact prediction of N=4 SUSYM theory for string theory,''
  J.\ Math.\ Phys.\  {\bf 42}, 2896 (2001)
  [hep-th/0010274].

\bibitem{esz} 
  J.~K.~Erickson, G.~W.~Semenoff and K.~Zarembo,
  ``Wilson loops in N=4 supersymmetric Yang-Mills theory,''
  Nucl.\ Phys.\ B {\bf 582}, 155 (2000)
  [hep-th/0003055].

\bibitem{Passerini:2011fe} 
  F.~Passerini and K.~Zarembo,
  ``Wilson Loops in N=2 Super-Yang-Mills from Matrix Model,''
  JHEP {\bf 1109}, 102 (2011)
  [JHEP {\bf 1110}, 065 (2011)]
  [arXiv:1106.5763 [hep-th]].

\bibitem{Fraser:2011qa} 
  B.~Fraser and S.~P.~Kumar,
  ``Large rank Wilson loops in N=2 superconformal QCD at strong coupling,''
  JHEP {\bf 1203}, 077 (2012)
  [arXiv:1112.5182 [hep-th]].


\bibitem{zaremborusso1} 
  J.~G.~Russo and K.~Zarembo,
  ``Large N Limit of N=2 SU(N) Gauge Theories from Localization,''
  JHEP {\bf 1210}, 082 (2012)
  [arXiv:1207.3806 [hep-th]].


\bibitem{zaremborusso2} 
  J.~G.~Russo and K.~Zarembo,
  ``Evidence for Large-N Phase Transitions in N=2* Theory,''
  JHEP {\bf 1304}, 065 (2013)
  [arXiv:1302.6968 [hep-th]].

\bibitem{zaremborusso3} 
  J.~G.~Russo and K.~Zarembo,
  ``Massive N=2 Gauge Theories at Large N,''
  JHEP {\bf 1311}, 130 (2013)
  [arXiv:1309.1004 [hep-th]].

\bibitem{Dorey:1999sj} 
  N.~Dorey,
  ``An Elliptic superpotential for softly broken N=4 supersymmetric Yang-Mills theory,''
  JHEP {\bf 9907}, 021 (1999)
  [hep-th/9906011].


\bibitem{Dorey:2002ad} 
  N.~Dorey and A.~Sinkovics,
  ``N=1* vacua, fuzzy spheres and integrable systems,''
  JHEP {\bf 0207}, 032 (2002)
  [hep-th/0205151].



\bibitem{mm} 
  N.~Dorey, T.~J.~Hollowood, S.~P.~Kumar and A.~Sinkovics,
  ``Exact superpotentials from matrix models,''
  JHEP {\bf 0211}, 039 (2002)
  [hep-th/0209089].






\bibitem{sw} 
  N.~Seiberg and E.~Witten,
  ``Electric - magnetic duality, monopole condensation, and confinement in N=2 supersymmetric Yang-Mills theory,''
  Nucl.\ Phys.\ B {\bf 426}, 19 (1994)
  [Erratum-ibid.\ B {\bf 430}, 485 (1994)]
  [hep-th/9407087].


\bibitem{sw2} 
  N.~Seiberg and E.~Witten,
  ``Monopoles, duality and chiral symmetry breaking in N=2 supersymmetric QCD,''
  Nucl.\ Phys.\ B {\bf 431}, 484 (1994)
  [hep-th/9408099].
  
  
\bibitem{dw} 
  R.~Donagi and E.~Witten,
  ``Supersymmetric Yang-Mills theory and integrable systems,''
  Nucl.\ Phys.\ B {\bf 460}, 299 (1996)
  [hep-th/9510101].


\bibitem{Dorey:2000fc} 
  N.~Dorey and S.~P.~Kumar,
  ``Softly broken N=4 supersymmetry in the large N limit,''
  JHEP {\bf 0002}, 006 (2000)
  [hep-th/0001103].


\bibitem{Aharony:2000nt} 
  O.~Aharony, N.~Dorey and S.~P.~Kumar,
  ``New modular invariance in the N=1* theory, operator mixings and supergravity singularities,''
  JHEP {\bf 0006}, 026 (2000)
  [hep-th/0006008].

\bibitem{Argyres:1995jj} 
  P.~C.~Argyres and M.~R.~Douglas,
  ``New phenomena in SU(3) supersymmetric gauge theory,''
  Nucl.\ Phys.\ B {\bf 448}, 93 (1995)
  [hep-th/9505062].

\bibitem{russo} 
  J.~G.~Russo,
  ``$ \mathcal{N} $ = 2 gauge theories and quantum phases,''
  JHEP {\bf 1412}, 169 (2014)
  [arXiv:1411.2602 [hep-th]].
  
\bibitem{Russo:2015vva} 
  J.~G.~Russo,
  ``Large $N_c$ from Seiberg-Witten Curve and Localization,''
  Phys.\ Lett.\ B {\bf 748}, 19 (2015)
  [arXiv:1504.02958 [hep-th]].

\bibitem{Nekrasov:2002qd} 
  N.~A.~Nekrasov,
``Seiberg-Witten prepotential from instanton counting,''
  Adv.\ Theor.\ Math.\ Phys.\  {\bf 7}, 831 (2004)
  [hep-th/0206161].

\bibitem{Nekrasov:2003rj} 
  N.~Nekrasov and A.~Okounkov,
  ``Seiberg-Witten theory and random partitions,''
  hep-th/0306238.

\bibitem{DV} 
  R.~Dijkgraaf and C.~Vafa,
  ``A Perturbative window into nonperturbative physics,''
  hep-th/0208048.



\bibitem{kkn} 
  V.~A.~Kazakov, I.~K.~Kostov and N.~A.~Nekrasov,
  ``D particles, matrix integrals and KP hierarchy,''
  Nucl.\ Phys.\ B {\bf 557}, 413 (1999)
  [hep-th/9810035].


\bibitem{Fraser:1996pw} 
  C.~Fraser and T.~J.~Hollowood,
  ``On the weak coupling spectrum of N=2 supersymmetric SU(n) gauge theory,''
  Nucl.\ Phys.\ B {\bf 490}, 217 (1997)
  [hep-th/9610142].



\bibitem{billo} 
  M.~Billo, M.~Frau, F.~Fucito, A.~Lerda, J.~F.~Morales, R.~Poghossian and D.~Ricci Pacifici,
  ``Modular anomaly equations in $ \mathcal{N} =2^*$ theories and their large-$N$ limit,''
  JHEP {\bf 1410}, 131 (2014)
  [arXiv:1406.7255 [hep-th]].

\bibitem{Ferrari:1996de} 
  F.~Ferrari,
  ``Charge fractionization in N=2 supersymmetric QCD,''
  Phys.\ Rev.\ Lett.\  {\bf 78}, 795 (1997)
  [hep-th/9609101].

\bibitem{Konishi:1998mk} 
  K.~Konishi and H.~Terao,
  ``CP, charge fractionalizations and low-energy effective actions in the SU(2) Seiberg-Witten theories with quarks,''
  Nucl.\ Phys.\ B {\bf 511}, 264 (1998)
  [hep-th/9707005].

\bibitem{dkm} 
  N.~Dorey, V.~V.~Khoze and M.~P.~Mattis,
  ``On mass deformed N=4 supersymmetric Yang-Mills theory,''
  Phys.\ Lett.\ B {\bf 396}, 141 (1997)
  [hep-th/9612231].
  



\bibitem{Minahan:1997if} 
  J.~A.~Minahan, D.~Nemeschansky and N.~P.~Warner,
  ``Instanton expansions for mass deformed N=4 superYang-Mills theories,''
  Nucl.\ Phys.\ B {\bf 528}, 109 (1998)
  [hep-th/9710146].
  
\bibitem{ritz} 
  A.~Ritz and A.~I.~Vainshtein,
  ``Long range forces and supersymmetric bound states,''
  Nucl.\ Phys.\ B {\bf 617}, 43 (2001)
  [hep-th/0102121].





\bibitem{Hollowood:2003cv} 
  T.~J.~Hollowood, A.~Iqbal and C.~Vafa,
  ``Matrix models, geometric engineering and elliptic genera,''
  JHEP {\bf 0803}, 069 (2008)
  [hep-th/0310272].

  


\bibitem{vw} 
  C.~Vafa and E.~Witten,
  ``A Strong coupling test of S duality,''
  Nucl.\ Phys.\ B {\bf 431}, 3 (1994)
  [hep-th/9408074].
  

\bibitem{zarembo} 
  K.~Zarembo,
  ``Strong-Coupling Phases of Planar N=2* Super-Yang-Mills Theory,''
  Theor.\ Math.\ Phys.\  {\bf 181}, no. 3, 1522 (2014)
  [arXiv:1410.6114 [hep-th]].





\bibitem{Martinec:1995by} 
  E.~J.~Martinec and N.~P.~Warner,
  ``Integrable systems and supersymmetric gauge theory,''
  Nucl.\ Phys.\ B {\bf 459}, 97 (1996)
  [hep-th/9509161].
  E.~J.~Martinec and N.~P.~Warner,
  ``Integrability in N=2 gauge theory: A Proof,''
  hep-th/9511052.
  
\bibitem{Gorsky:1995zq} 
  A.~Gorsky, I.~Krichever, A.~Marshakov, A.~Mironov and A.~Morozov,
  ``Integrability and Seiberg-Witten exact solution,''
  Phys.\ Lett.\ B {\bf 355}, 466 (1995)
  [hep-th/9505035].



 
 
\bibitem{Douglas:1995nw} 
  M.~R.~Douglas and S.~H.~Shenker,
  ``Dynamics of SU(N) supersymmetric gauge theory,''
  Nucl.\ Phys.\ B {\bf 447}, 271 (1995)
  [hep-th/9503163].

  


  

  
\bibitem{ww} E. T. Whittaker and G. N. Watson, `` A Course of Modern Analysis,'' Cambridge University Press, 1927.

\bibitem{koblitz} N. Koblitz, `` Introduction to Elliptic Curves and Modular Forms ,'' Springer-Verlag, 1984.



  





  
  
  
  

\end{thebibliography}
\end{document}